\documentclass[a4paper,11pt]{article}

\usepackage{graphicx,amssymb,amsmath,color}
\usepackage[left=2.5cm,right=2.5cm,top=3cm,bottom=3cm]{geometry}

\usepackage[normalem]{ulem}
	
\newcommand{\moire}{moir\'e }
\newcommand{\degree}{$^\circ$ }

\newcounter{bibnum}

\title{{Momentum-space} indirect interlayer excitons in {transition metal dichalcogenide} van der Waals heterostructures}

\author{
Jens Kunstmann,$^{1\dag\S}$
Fabian Mooshammer,$^{2\dag}$
Philipp Nagler,$^{2}$
Andrey Chaves,$^{3,4}$ \\
Frederick Stein,$^{1}$
Nicola Paradiso,$^{2}$
Gerd Plechinger,$^{2}$
Christoph Strunk,$^{2}$\\
Christian Sch\"uller,$^{2}$
Gotthard Seifert,$^{1,5}$
David R. Reichman,$^{4}$
Tobias Korn$^{2\S}$
\thanks{
$^1$ Theoretical Chemistry, Department of Chemistry and Food Chemistry, TU Dresden, 01062 Dresden, Germany \newline
$^2$ Institut f\"ur Experimentelle und Angewandte Physik, Universit\"at Regensburg, D-93040 Regensburg, Germany \newline
$^3$ Departamento de Fisica, Universidade Federal do Ceara, Caixa Postal 6030, Campus do Pici, 60455-900 Fortaleza, Ceara, Brazil \newline
$^4$ Department of Chemistry, Columbia University, New York, NY 10027, USA\newline
{$^5$ National University of Science and Technology, MISIS, 119049 Moscow, Russia}\newline	
$\dag$ These authors contributed equally \newline
$\S$ Correspondence should be addressed to J.K.~(e-mail: jens.kunstmann@tu-dresden.de) or T.K.~(e-mail: tobias.korn@physik.uni-regensburg.de).
}
}

\date{}

\begin{document}

\maketitle


\textbf{
{Monolayers of transition metal dichalcogenides (TMDCs) feature exceptional optical properties that are dominated by excitons, tightly bound electron-hole pairs. Forming van der Waals heterostructures by deterministically stacking individual monolayers allows to tune various properties via choice of materials \cite{Geim2013} and relative orientation of the layers \cite{VanderZande2014,Liu2014h}.}
In these structures, a new type of exciton emerges, where electron and hole are spatially separated. 
These interlayer excitons \cite{Fang2014d,Rivera2014,Rivera2016} allow exploration of many-body quantum phenomena \cite{Snoke2002, Fogler2014} and are ideally suited for valleytronic applications \cite{Xu2014}. Mostly, a basic model of fully spatially-separated electron and hole stemming from the $K$ valleys of the monolayer Brillouin zones is applied to describe such excitons. Here, we combine photoluminescence spectroscopy and first principle calculations to expand the concept of interlayer excitons. We identify a partially charge-separated electron-hole pair in MoS$_2$/WSe$_2$  heterostructures residing at the $\Gamma$ and $K$ valleys. We control the emission energy of this new type of {momentum}-space indirect, yet strongly-bound exciton by variation of the relative {orientation of the layers}. These findings represent a crucial step towards the understanding and control of excitonic effects in TMDC heterostructures
and devices.
}

\begin{figure*}
\centering
\includegraphics[width=.5\textwidth]{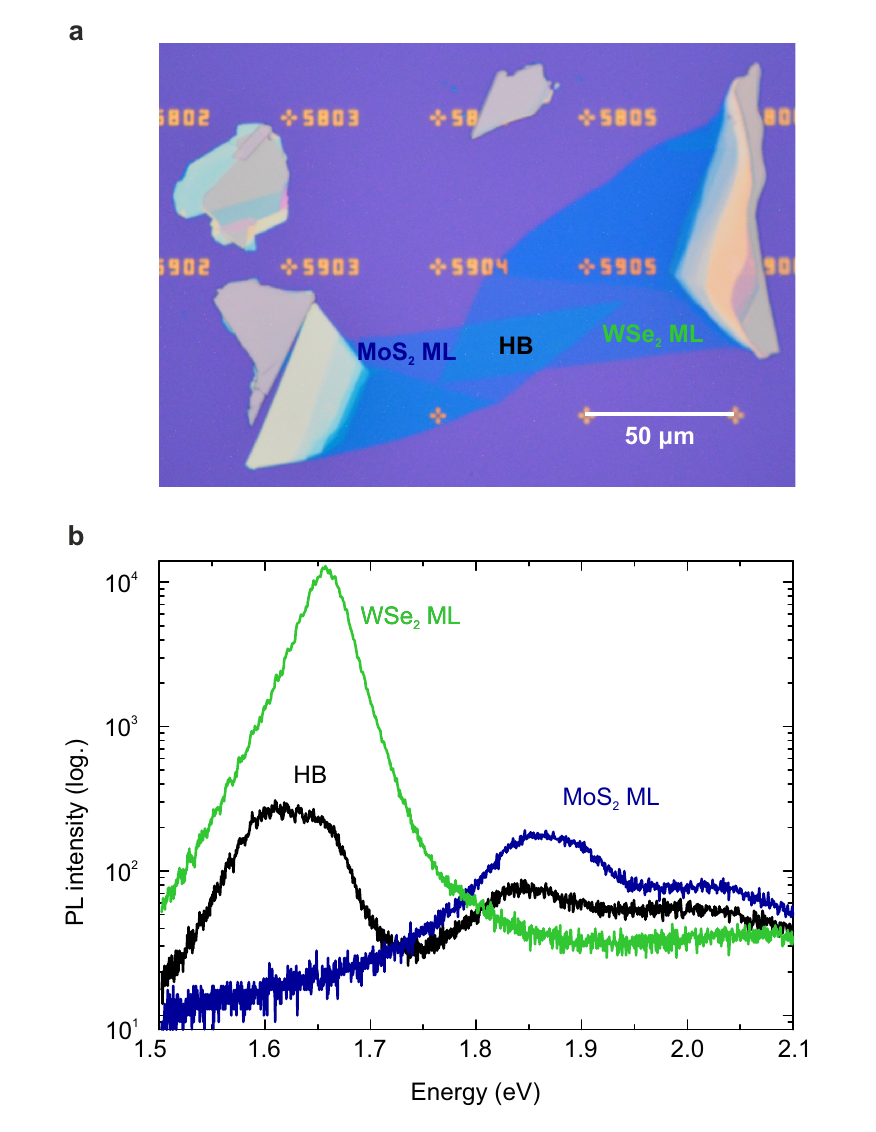}
\caption{\label{fig:system}
\textbf{{Interlayer excitons in MoS$_2$/WSe$_2$ heterobilayers.}}
\textbf{a}, Optical micrograph of a sample with a twist angle of $ 58.7 ^{\circ}$, fabricated by deterministic transfer and stacking. Monolayer (ML) and heterobilayer (HB) regions are indicated.
\textbf{b}, Photoluminescence (PL) spectra of the HB and the ML regions. The occurrence of an interlayer exciton (ILE) near 1.6 eV is discernible in the HB.
}
\end{figure*}

\begin{figure*}
\centering
\includegraphics[width=.92\textwidth]{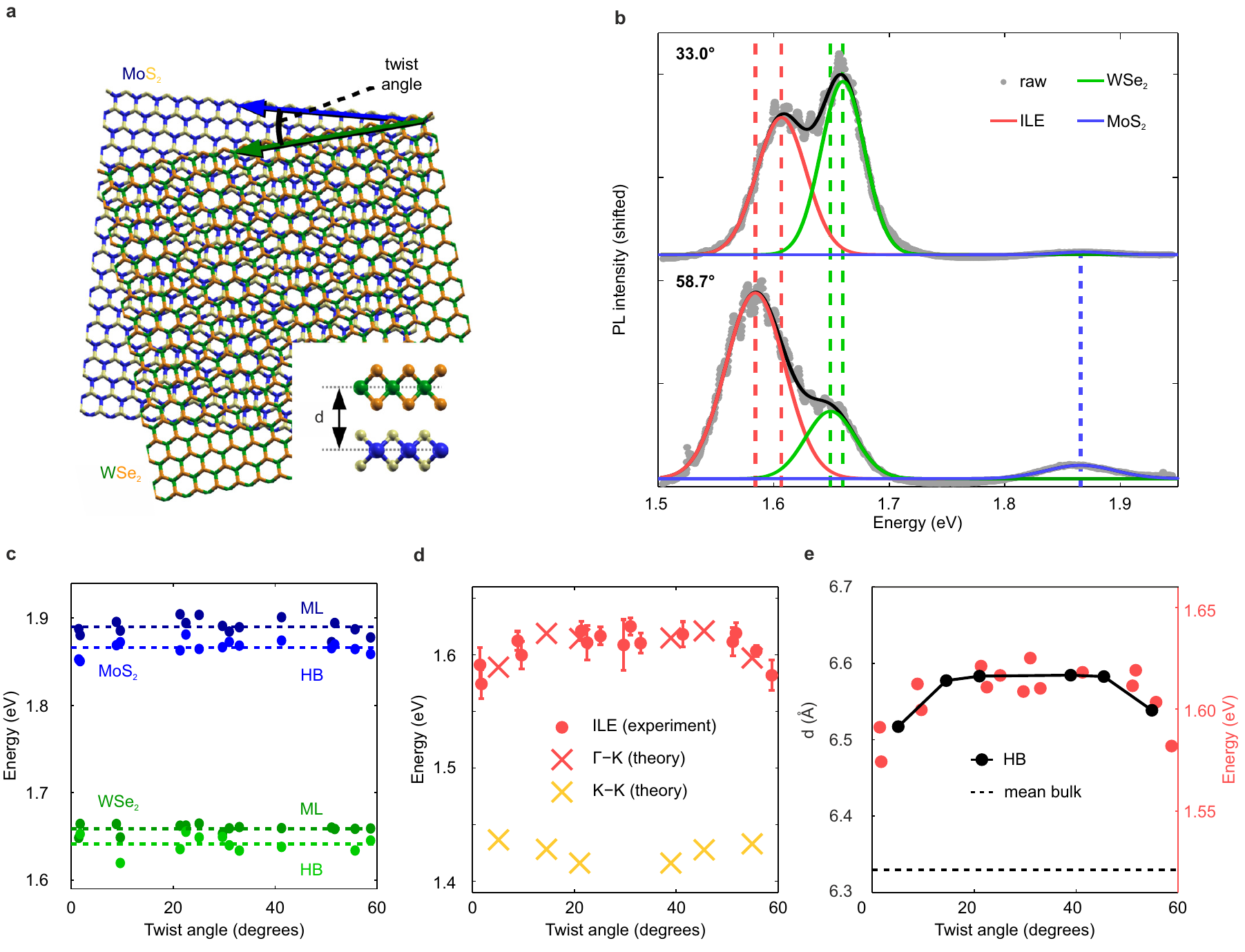}
\caption{\label{fig:PL}	
\textbf{{Tuning the interlayer exciton energy via interlayer twist.}}
\textbf{a}, Atomic structure illustration of MoS$_2$/WSe$_2$ HB. The twist angle is the relative lattice orientation of the two layers.
Inset: side view of the HB; the layer separation $d$ is the distance between Mo and W planes.
\textbf{b}, PL spectra and their decomposition into Gaussian peaks for two HB with twist angles of 33.0\degree and 58.7\degree. Besides the A exciton peaks, arising from the WSe$_2$ (green) and MoS$_2$ (blue), an ILE near 1.6 eV has emerged (red). Dashed vertical lines allow to compare peak positions; black lines are the sum of the Gaussians.
The extracted peak energies are used in \textbf{c} and \textbf{d}. 	
\textbf{c}, A exciton energies for ML and HB regions for varying twist angles. Dashed horizontal lines indicate the mean values. We observe no clear dependence on the twist angle but a redshift from the monolayer to the HB.
\textbf{d}, ILE energies and calculated transition energies for HB with different twist angles. {The error bars indicate the standard deviation of the ILE energy determined from spatial averaging of ILE PL emission (see Supplementary Information).} The $\Gamma-K$ and $K-K$ values are calculated with density functional theory and they are rigidly upshifted by 0.445 eV (see text). Only the trend of $\Gamma-K$ is in quantitative agreement with the experiment.
\textbf{e}, Mean layer separation (indicated graphically in the inset of \textbf{a}) as function of twist angle, as calculated with dispersion-corrected density functional theory.
Steric repulsion of chalcogen atoms, due to lattice mismatch and incommensurability, creates a twist angle dependence and leads to bigger layer separations than the mean of the layer spacings of bulk MoS$_2$ and WSe$_2$ samples (dashed horizontal line). Red dots correspond to 'ILE (experiment)' in \textbf{d}. A strong correlation is discernible; the linear proportionality factor is 0.44 eV/{\AA}.
}
\end{figure*}

{
An optical micrograph of a representative MoS$_2$/WSe$_2$ heterobilayer (HB), which was fabricated by deterministic transfer and stacking~\cite{Castellanos2014} followed by an annealing procedure, is shown in Fig.~\ref{fig:system}(a).
All HB and isolated regions of the constituent monolayers (ML) were thoroughly studied by micro-photoluminescence (PL) spectroscopy, typical spectra are shown in  Fig.~\ref{fig:system}(b).
The ML regions display the well-known A exciton and trion peaks \cite{Splendiani2010,Mak2010,Zhao2013d,Mak2013} near 1.65 and 1.9 eV for WSe$_2$ (green) and MoS$_2$ (blue), respectively.
In the HB region the same two peaks are discernible, but slightly shifted in energy due to the modified dielectric environment~\cite{Latini2017,Raja2017}.
However, in addition a new peak near 1.6 eV is observed, which is absent in the  ML regions.
We assign this peak to the interlayer exciton (ILE) \cite{Fang2014d,Chiu2014}.

Now, we control the relative orientation of the TMDC layers to reveal the $k$-space indirect nature of this ILE in MoS$_2$/WSe$_2$ HBs.
The twist angle
is measured with respect to each layer's zigzag direction (green and blue arrows in Fig.~\ref{fig:PL}(a)), varying between 0\degree (aligned) and 60\degree (antialigned).
A total of 15 HBs with twist angles covering this range were fabricated, and the ILE emission was observable as a high-intensity PL peak in all samples.
The twist angle was determined by second harmonic generation measurements and the samples were further characterized by Raman spectroscopy (for details see Supplementary Information).}
The presence of the ILE becomes more obvious in Figure \ref{fig:PL}(b), which displays PL spectra from the HB region of two samples with twist angles of 33.0\degree and 58.7\degree and their decomposition into three Gaussian peaks.
The comparison of the two panels shows that as the twist angle is varied, the ILE peak shifts much more in energy than the A/trion peaks.
As is clear from Fig.~\ref{fig:PL}(c), the latter do not exhibit a distinct dependence on the twist angle.
Figure \ref{fig:PL}(d) shows that the ILE energy (red balls) shifts as a continuous function of the twist angle over a range of 50 meV. The maximum of the curve is near 30\degree and it exhibits a slight asymmetry, i.e., the minimum near 0\degree has a smaller energy than the minimum near 60\degree. Similar twist-angle-dependent, slightly asymmetric shifts of PL peaks have been observed in twisted bilayer MoS$_2$~\cite{VanderZande2014,Liu2014h,Yeh2016a}.

\begin{figure*}
\includegraphics[width=\textwidth]{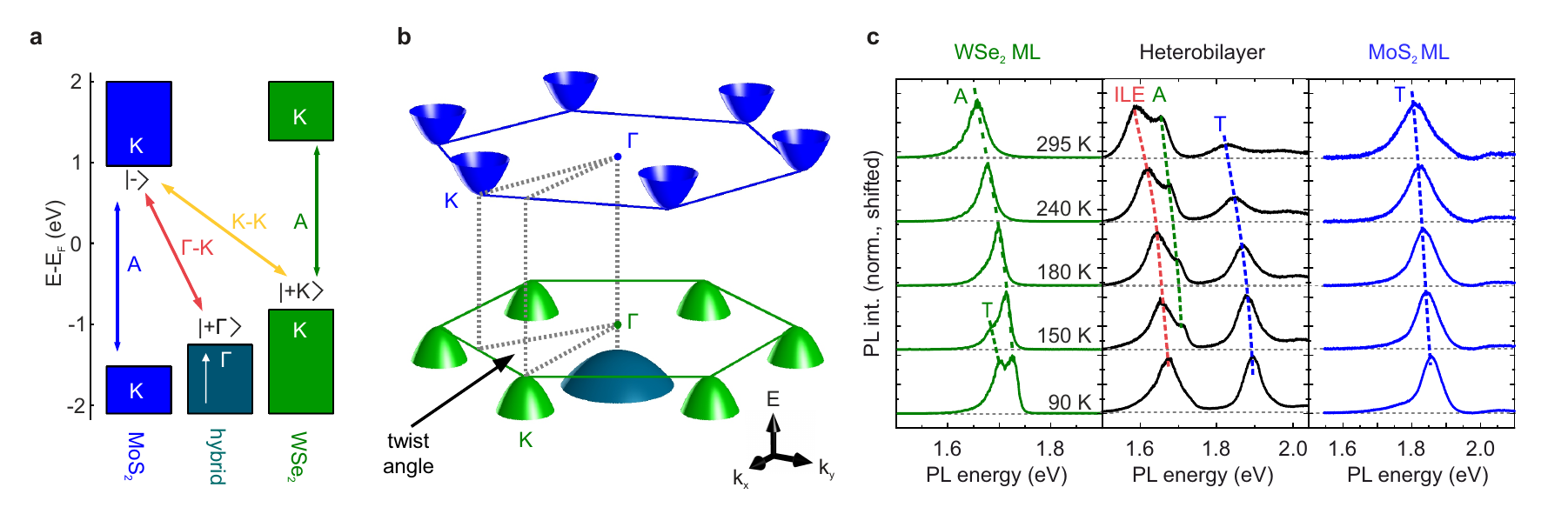}
\caption{\label{fig:ElecStr}	
\textbf{Electronic structure of MoS$_2$/WSe$_2$ heterobilayers.}
\textbf{a}, Band alignment diagram. The valence band maximum at the $\Gamma$ valley $|+\Gamma\rangle$ is a hybrid state of both layers and it moves up as hybridization increases.
Colored vertical arrows  indicate ML transitions. $\Gamma-K$ and $K-K$ are possible interlayer transitions.
\textbf{b}, Two-dimensional band structure of the HB near the band edges. In twisted HB the  Brillouin zones of MoS$_2$ (blue) and WSe$_2$ (green) are misaligned. Therefore both HB transitions, $K-K$ and $\Gamma-K$, are $k$-space indirect {(the wave vectors of the electron and holes states differ)}. However,  {for twist angles near 0\degree (aligned) or 60\degree (antialigned)} $K-K$ is $k$-space direct {(no wave vector difference)}.
\textbf{c}, Temperature-dependent PL spectra measured on isolated  WSe$_2$ (left panel)  and MoS$_2$ (right panel)  MLs and on a  HB region (center panel). {Every spectrum is individually normalized to the peak of highest intensity. In the HB region, spatial averaging of PL spectra is performed due to the spatially inhomogeneous ILE emission, see Supplementary Information for details. The dotted lines trace the spectral evolution of the WSe$_2$ and MoS$_2$ A exciton and trion, as well as the ILE as a function of temperature. As the temperature is decreased, the ILE PL, which is the most prominent emission at room temperature, is suppressed compared to the \emph{intralayer} MoS$_2$ emission in the HB region. This finding supports our assertion that the ILE is related to a $k$-space-indirect, phonon-assisted transition.}
}
\end{figure*}

We are able to explain this effect quantitatively via density functional theory (DFT) calculations. Details may be found in the Supplementary Information.  An analysis of the geometries revealed that the mean layer separation of a HB changes as a continuous function of the twist angle over a range of 0.07 {\AA}, as shown in Fig.~\ref{fig:PL}(e).  This result can be ascribed to steric effects since the surface of a TMDC ML is not atomically smooth but corrugated due to  protrusion of the chalcogen atoms out of the metal atom plane.
For angles near 0\degree or 60\degree, Fig.~\ref{fig:PL}(e) indicates a reduction of the mean layer separation by 1\%. In these systems long-wavelength \moire patterns are formed and the individual layers maximize their adhesion via adopting static spatial fluctuations.
To study the consequences of these observations we now analyze the electronic structure of the MoS$_2$/WSe$_2$ HB (for details see Supplementary Information). A level alignment diagram is shown in Fig.~\ref{fig:ElecStr}(a). It illustrates the staggered band alignment of the HB, the optical transitions in the two MLs that give rise to the A excitons (vertical arrows) and the $K-K$ and $\Gamma-K$ interlayer transitions. {Due to the generally weak coupling between the MLs in TMDC heterostructures the Bloch wave vectors defining the $K$ valleys of the MLs are also approximately good quantum numbers of the HB.}
In Fig.~\ref{fig:PL}(d) $K-K$ and $\Gamma-K$ interlayer transition energies of twisted HB, as obtained from  DFT calculations, are plotted as function of twist angle. To allow a visual comparison of the DFT transition energies with the PL ILE energies, the DFT values in Fig.~\ref{fig:PL}(d) are rigidly shifted by 0.445 eV, {which implies that (relative) energy differences and not absolute energies are compared.}
The comparison reveals a remarkable quantitative correspondence with the $\Gamma-K$ transition (red) but not with $K-K$ (yellow). This suggests that the ILE is related to the $\Gamma-K$ transition.
For the ML A transitions the DFT results exhibit no change with twist angle (see Supplementary Information).
{The behavior exposed by the DFT calculations is in full agreement with the PL results of Fig.~\ref{fig:PL}(d), because the  change of the $\Gamma-K$ transition energy is essentially a shift of the $\Gamma$-point valence band energy (white arrow in Fig.~{3(a)}), an effect that should be well captured by DFT (for details see Supplementary Information Sec.~3.2.4) \cite{Yeh2016a,Zhu1989a}.}

To better understand the impact of the layer separation on the electronic structure of the MoS$_2$/WSe$_2$ HB, we studied by DFT an artificial model system that is antialigned and lattice-commensurate by applying strain.
We considered 28 different transitions between valence and conduction band extrema and calculated their energies as function of layer separation (see Supplementary Information). Most transitions exhibit either no dependence on the layer separation or a linear dependence with a negative slope (including $K-K$).
There is only a single transition, $\Gamma-K$, that lies within a reasonable energy range, has the correct trend and a positive slope of 0.47 eV/{\AA}, in excellent agreement with 0.44 eV/{\AA} found for realistic systems in Fig.~\ref{fig:PL}(e).
These results uniquely identify the {observed} ILE to be related to the $\Gamma-K$ transition and  {not to} $K-K$  that is usually assumed when studying ILEs. {Additional evidence supporting this key finding of our work is provided by analysing the twist angle and temperature dependencies of the ILE PL intensity and by an exciton model.}

We note that if the ILE was related to a $K-K$ transition, its PL emission should only be observable for nearly (anti-)aligned structures because the transition probability of $k$-space direct transitions is higher~\cite{Nayak2017}. However, the analysis of the PL intensity as a function of twist angle {in the HBs} shows no pronounced angle-dependence (see Supplementary Information).

For indirect optical transitions the difference between the wave vectors of the electron and holes states is compensated by coupling to a phonon and {the efficiency of this process} can be {partially tuned by varying the temperature, which controls the phonon population}.
Temperature-dependent PL measurements of a HB and isolated WSe$_2$ and MoS$_2$ MLs are shown in Fig.~\ref{fig:ElecStr}(c). We observe a systematic blueshift of all exciton peaks with decreasing temperature.
{Additionally, we observe a complex behavior of the relative PL intensities in the heterobilayer region. The ILE PL, which is the most prominent emission peak at room temperature, decreases relative to the MoS$_2$ \emph{intralayer} emission as the temperature decreases, in stark contrast to the behavior for ILE of the $K-K$ type observed in WSe$_2$-MoS$_2$ heterobilayers, where the ILE PL yield monotonously increases with decreasing temperature~\cite{Rivera2014,Nagler17}, further supporting  the $k$-space indirect character of the transition in our heterobilayer system. We also observe that the WSe$_2$ \emph{intralayer} emission is quenched with decreasing temperature, as reported previously for  WSe$_2$ monolayers~\cite{Zhang:2015d,Arora:2015b}. Additional measurements and discussion of the temperature-dependent PL are presented in the Supplementary Information}.

\begin{figure*}
\includegraphics[width=1.03\textwidth]{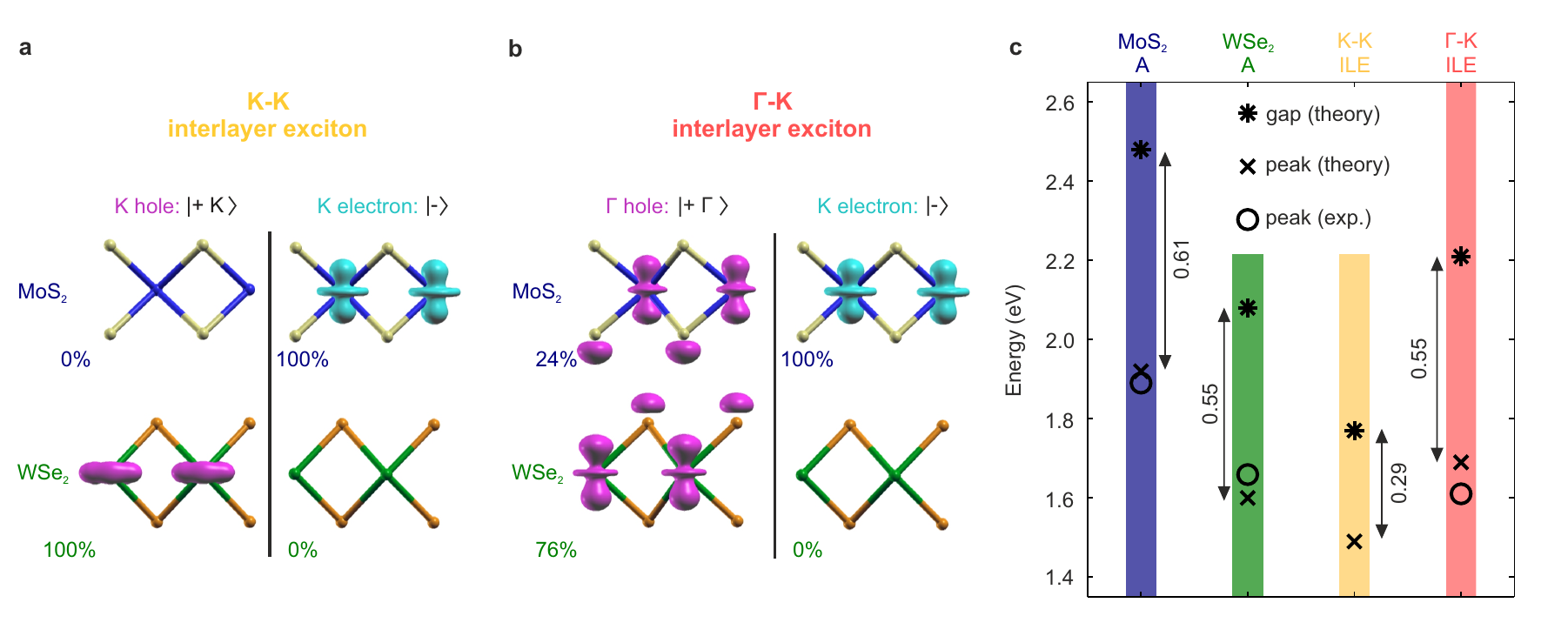}
\caption{\label{fig:excitons}
\textbf{The nature of interlayer {excitons}.}
\textbf{a}, The hole $|+K\rangle$ and electron $|-\rangle$ states of the $K-K $ ILE are localized in the individual layers (the pink and magenta contours are partial charge densities).
\textbf{b}, Hole and electron states of the $\Gamma-K$ ILE. While the electron state $|-\rangle$ is only localized in the MoS$_2$ layer, the hybrid hole $|+\Gamma\rangle$ state is delocalized over both layers. Values in percent correspond to the fraction of the wave function that is localized in each layer.
\textbf{c}, Comparison between experimental ($\circ$) and theoretical ($\times$) photoluminescence peak energies for different monolayer A excitons and ILE. The arrows indicate exciton binding energies $\Delta E_X$, as calculated with our model. The theoretical band gaps ($\ast$) are G$_0$W$_0$ results from the literature \cite{Rasmussen2015}. The theoretical results {closely match  the experimental data}.
$\Delta E_X$ for the $\Gamma-K$ ILE is comparable to values in the ML, {and the resulting emission energy is in good agreement with the experimentally observed ILE peak}.
}
\end{figure*}

We now analyze the localization of electron and hole wave functions in the WSe$_2$/MoS$_2$ HB.
Figures \ref{fig:excitons}(a),(b) show partial charge densities of electron and hole states for the $\Gamma-K$  and $K-K$ transitions in the model system calculated with DFT.
Three unique states are involved, the $K$-valley electron state $|-\rangle$, the $K$-valley hole state $|+K\rangle$ and the $\Gamma$-valley hole state $|+\Gamma\rangle$.
The electron-hole wave function overlap of a ILE can be quantified by projecting the  hole state $|+ k \rangle$ ($k=\Gamma$ or $K$) onto the MoS$_2$ layer $o_k = |\langle \text{MoS}_2|+ k \rangle|^2$. The overlap of the $K-K$ transition (see Fig.~\ref{fig:excitons}(a)) is nearly zero ($o_K \approx 0$ \%) because electrons and holes
only involve transition metal atom d-states and
reside 6.6~{\AA} apart (see Fig.~\ref{fig:PL}(e)).
{The PL intensity scales with the square of the transition matrix element, which}
suggests that radiative recombination of $K-K$ ILE is suppressed and is thus not seen in our PL measurements \cite{Kira1998}.
This is very different for the $\Gamma-K$ transition (Fig.~\ref{fig:excitons}(b)):
$|+\Gamma\rangle$ is strongly affected by interlayer hybridization and therefore extends over both layers with Mo, S, W, and Se atoms all participating.
It has a large component that resides in the MoS$_2$ ML ($o_\Gamma = 24$\%) where $|-\rangle$ is localized. Therefore {the matrix element} is much larger for $\Gamma-K$ transitions than it is for $K-K$ ones.

Our observations strongly imply the picture of an ILE {with high PL intensity that} does not represent the thermodynamically lowest-energy states (i.e. the $K-K$ excitonic transition),  {and} is fully consistent with transitions of the  $\Gamma-K$ type. We note that our system is pumped with a sufficiently high energy to create carriers across a wide range of momenta with a hot, non-thermal distribution.  Thus the observed response will depend intimately on the non-equilibrium kinetics of exciton formation and recombination, {as well as charge transfer~\cite{Hong2014} and a host of  non-radiative relaxation channels \cite{Sun14,Steinleitner2017}}. However, it should be noted that non-equilibrium effects alone are insufficient to explain why the $\Gamma-K$ ILE appears to be so strongly favored.   One possibility is that the large, real-space overlap of electron and holes in the respective layers kinetically favors the formation and recombination of partially charge-separated $\Gamma-
K$ excitons despite the fact that such
states are not formed from band edge carriers.
{The large, real-space overlap of these k-space indirect interlayer excitons suggests a binding energy that is increased as compared to their $K-K$ counterparts that are fully charge-separated. Therefore, we calculate the exciton binding energies} $\Delta E_X$ of the A excitons as well as $K-K$ and $\Gamma-K$ ILEs using the Quantum Electrostatic Heterostructure model \cite{Andersen2015} and a variational wave function ansatz. Excitonic interlayer  interactions are described within a tight-binding approach (see Supplementary Information for details).
Experimentally{,} the exciton binding energy is defined as $\Delta E_X = E_\mathrm{gap}^\mathrm{qp} - E_\mathrm{gap}^\mathrm{opt}$, where $E_\mathrm{gap}^\mathrm{qp}$ is the quasiparticle band gap and $E_\mathrm{gap}^\mathrm{opt}$ is the optical gap, measured as the PL peak energy.
The results of these calculations {are} given in Fig.~\ref{fig:excitons}(c), where $\Delta E_X$ is indicated by arrows.
For the A excitons{,} the theoretical and experimental energies agree well. $\Delta E_X$ values are of the order of 0.5 eV, in good agreement with previous results~\cite{Rasmussen2015}. For the $K-K$ ILE $\Delta E_X = 0.29$ eV, which is also in agreement with earlier results \cite{Latini2017,Wilson2017}.
For $\Gamma-K$ we obtain a much bigger value of 0.55 eV, comparable to those of A excitons. The main reason for this large number is the delocalization of the hole state over both layers that enhances the electron-hole Coulomb attraction and gives the $\Gamma-K$ ILE also a strong ML character in MoS$_2$. {We note that the ILE emission energy calculated using this large binding energy is in good agreement with the experimentally observed value.}
It should also be noted that charge separation creates excitons with an interlayer dipole moment of  $\mu_\mathrm{IL} = (1-o_k) e d \approx (1-o_k) \cdot 1.4$ Debye ($e$ is the elementary charge and $d$ the layer separation). The latter is reduced by interlayer hybridization. Thus $K-K$ and $\Gamma-K$ excitons can potentially be distinguished by measuring $\mu_\mathrm{IL}$.

In this work we have shown that MoS$_2$/WSe$_2$ HBs host optically bright, $k$-space indirect excitons composed of holes from the $\Gamma$-valley and electrons from the $K$-valley. The  $\Gamma-K$ character of these interlayer excitons  was uniquely identified by the twist angle dependence of the PL emission energy in conjunction with first principles calculations that quantitatively reproduce the observed energy shifts. Their $k$-space indirect nature is further supported by analysis of temperature- and twist-angle-dependent PL intensities, and by a $\Gamma-K$ exciton model yielding the emission energy observed in experiment. We further showed that strong interlayer hybridization of the hole state reduces the degree of charge separation and enhances radiative recombination that gives rise to the high PL intensity of the ILE, while preserving an exciton binding energy comparable to monolayer excitons. These results extend current interpretations about the nature of ILE in TMDC-based van der Waals 
heterostructures.
{with important implications for both the study of fundamental physics as well as the development of optoelectronic devices.}

\subsection*{Methods}
See {the Supplementary Information in} the on-line version of the paper.



\subsection*{Acknowledgements}
The work is financially supported by the German Research Foundation (DFG) under grant numbers SE 651/45-1, GRK 1570, KO 3612/1-1 and KO 3612/3-1.
G.S.~gratefully acknowledges financial support by the Ministry of Education and Science of the Russian Federation (No. K3-2017-064). 
Computational resources for this project were provided by ZIH Dresden.

\subsection*{Author contributions}
F.M., P.N., C.S. and T.K. conceived the experiments. F.M. fabricated the samples and performed the optical spectroscopy and data analysis together with P.N. and T.K.
J.K did the DFT calculations together with F.S and G.S., interpreted the results and supervised the theoretical analysis.  A.C. did the exciton modeling under supervision of D.R.R. using parameters provided by J.K.
J.K, D.R.R. and T.K. wrote the paper.
All authors discussed the results.

\subsection*{Additional information}
Supplementary Information is available in the on-line version of the paper.
{
The data that support the plots within this paper and other findings of this study are available from the corresponding author upon reasonable request.
}

\subsection*{Competing financial interests}
The authors declare no competing financial interests.

\newpage

%

\clearpage
\newpage

\subsection*{Methods (on-line)}
\subsubsection*{Sample fabrication}
Heterobilayer (HB) samples were fabricated by means of a deterministic transfer process~\cite{Castellanos2014}. For this, we initially exfoliated MoS$_2$ and  WSe$_2$ flakes from bulk crystals (here, we utilized a natural MoS$_2$ crystal and a synthetic WSe$_2$ crystal bought at HQgraphene.com) onto polydimethylsiloxane (PDMS) substrates. Monolayer (ML) regions of these flakes were identified via optical microscopy. Then, we first transferred a MoS$_2$ flake onto the target substrate, a silicon wafer piece covered with SiO$_2$ layer and pre-defined metal markers. Subsequently, the WSe$_2$ flake was transferred on top of the MoS$_2$. For each of the heterobilayers fabricated in this way, the relative orientation of the individual layers was chosen to optimize the overlap region of the monolayer parts of the flakes.  Subsequent to the transfer, the heterobilayers were annealed. For this, they were mounted in a furnace, which was initially flushed with an H$_2$/Argon gas {mixture} and then pumped to high
vacuum.
In vacuum, the samples were annealed at a temperature of 150$^o$C for 5 hours.

\subsubsection*{Optical spectroscopy}
Photoluminescence (PL) and Raman measurements were performed in a self-built microscope setup, details are published elsewhere~\cite{Plechinger2015_2Dmat}.
A continuous-wave laser (wavelength 532 nm) was coupled
{to} a 100x microscope objective and focussed to a sub-micron spot on the sample surface. PL and scattered light were collected with the same objective, passed through long-pass filters, coupled into a grating spectrometer and detected with a {P}eltier-cooled charge-coupled device (CCD). For temperature-dependent PL measurements, the sample was mounted on the cold finger of a small He-flow cryostat. For PL mapping, the sample was moved beneath the fixed microscope objective using a motorized xy stage, and PL spectra were collected for sample positions defined on a square lattice. In order to extract information from these spectra, an automated fitting routine was employed, which yields the integrated intensity, spectral position and full width at half maximum (FWHM) for each spectral feature extracted using a Gaussian fit function.
Second harmonic generation (SHG) measurements were performed in a similar, self-built microscope setup, which was optimized to yield high SHG throughput. Here,   a Ti:sapphire laser oscillator (pulse length 100~fs, central wavelength 815~nm) was used as an excitation source. The laser light coupled into the microscope objective was linearly polarized, and the same polarizer was used to analyze the reflected light, so that only the signal polarized parallel to the excitation was detected.  To separate the SHG signal from the reflected fundamental laser wavelength, a dichroic mirror and  short pass filters were employed before the SHG signal was either coupled into a grating spectrometer to be detected by a CCD, or focussed onto an avalanche photodiode (APD). In the measurements using the APD, a lock-in scheme was employed to further increase signal-to-noise ratio. For SHG mapping, the sample was moved beneath the fixed microscope objective using a motorized xy stage. For polarization-dependent measurements,
the combined polarizer/
analyzer was rotated using a motorized stage.

\subsubsection*{Experimental data analysis}
For each heterobilayer structure, a PL map was measured at room temperature. To compensate for the spatial inhomogeneity of the interlayer exciton (ILE) emission, {spatial averaging was employed. For this, the average PL emission energy of the ILE, as well as its standard deviation, were calculated from the values extracted from an automated fitting routine applied to individual PL spectra collected in the  heterobilayer region where sufficiently intense ILE PL was observed. The size of these regions varied from sample to sample, but on average, more than 60 spectra were evaluated for an individual heterobilayer.}

\subsubsection*{Density functional theory calculations}
Density functional theory (DFT) calculations were
carried out with the PBE functional \cite{Perdew1996} and the DFT-TS dispersion-interaction-correction scheme \cite{Tkatchenko2009} using the PAW method \cite{Blochl1994a} and a plane wave basis set with a cutoff energy of 259 eV, as implemented in the VASP package \cite{Kresse1996a,Kresse1996}. For the k-point sampling, an in-plane sampling density of 0.1 \AA$^{2}$ was used. It was carefully checked that this density leads to converged total energies (energy differences are smaller than 1 meV/atom). The k-space integration was carried out with a Gaussian smearing method using an energy width of 0.05 eV for all calculations. All unit cells were built with at least 10 {\AA} separation between replicas in the perpendicular direction to achieve negligible interaction. All systems were fully structurally optimized until all interatomic forces and stresses on the unit cell were below 0.01 eV/{\AA} and 10
kbar, respectively.
Spin-orbit interactions were generally not taken into account
and the inclusion of these interactions does not alter any of our conclusions as spin-orbit-dependent interlayer interactions have never been reported before \cite{VanderZande2014,Yeh2016a}.
The wave function overlap $o_k = |\langle \text{MoS}_2|+ k \rangle|^2$ is calculated by integrating the partial charge density of the state $|+ k \rangle$ ($k=\Gamma$ or $K$) over the half of the  volume of the unit cell that contains the MoS$_2$ layer. The cutting plane between the two halves is the minimum of the plane-averaged line charge density in the van der Waals gap between the layers.


\clearpage
\begin{center}
\textbf{\Large Supplementary Information for the article:\\
Momentum-space indirect interlayer excitons in transition metal dichalcogenide van der Waals heterostructures}
\end{center}
\setcounter{equation}{0}
\setcounter{figure}{0}
\setcounter{table}{0}
\setcounter{page}{1}
\renewcommand{\thetable}{S\arabic{table}}
\renewcommand{\thefigure}{S\arabic{figure}}
\tableofcontents

\section{Supporting experimental details}
\subsection{Evaluating the effects of annealing}
\begin{figure}
\centering
\includegraphics[width=0.7 \textwidth]{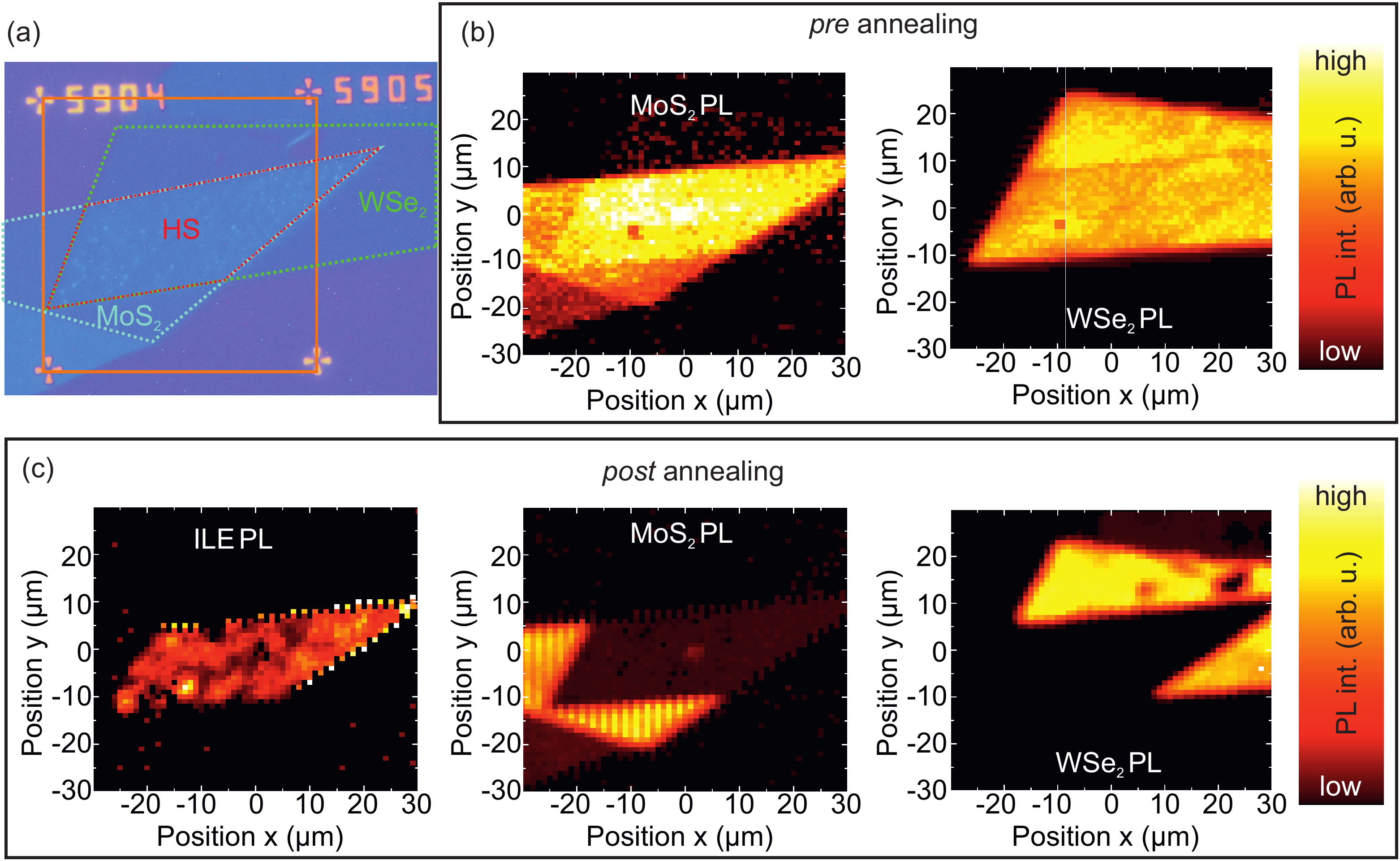}
\caption{\label{fig:Annealing_PL}(a) Optical microscope image of heterobilayer sample. The orange rectangle marks the  area of the PL scans shown in (b) and (c). The outlines of the MoS$_2$ and WSe$_2$ monolayers and the heterobilayer are marked by the dotted lines. (b) False color maps of the PL intensities of MoS$_2$ and WSe$_2$ emission, measured at room temperature before annealing of the sample. (c) False color maps of the PL intensities of interlayer exciton, MoS$_2$ and WSe$_2$ emission, measured at room temperature after annealing of the sample.
}
\end{figure}

All samples investigated
{in} this study were subjected to an annealing step after fabrication, which was previously shown to increase interlayer coupling~\cite{Tongay2014_SI}. Here, we demonstrate the effects of the annealing using photoluminescence measurements and atomic force microscopy. Figure~\ref{fig:Annealing_PL}(a) shows an optical microscope image of a specific heterobilayer sample. To evaluate the coupling between the two constituent monolayers in this sample, we performed scanning photoluminescence measurements before and after the annealing step. In Fig.~\ref{fig:Annealing_PL}(b), false-color maps of the PL intensities of MoS$_2$ and  WSe$_2$, measured before the annealing, are shown. We clearly observe that the PL intensity of both materials is not diminished in the heterobilayer region, {indeed} it is even slightly enhanced for MoS$_2$. This indicates that the two monolayers are decoupled, so that there is {neither} band structure hybridization {nor} rapid interlayer charge transfer. By contrast, after annealing,
we find a pronounced quenching of the PL emission from the MoS$_2$ and  WSe$_2$ in the heterobilayer region, while the emission is unchanged in the isolated monolayer regions, as Fig.~\ref{fig:Annealing_PL}(c) shows. Additionally, we observe the lower-energy PL emission of the interlayer exciton in the heterobilayer. This indicates that the annealing procedure enables electronic interlayer coupling and rapid charge transfer. We note that the interlayer exciton PL intensity is spatially inhomogeneous, indicating local variations of interlayer coupling.

\begin{figure}
\centering
\includegraphics[width= 0.7\textwidth]{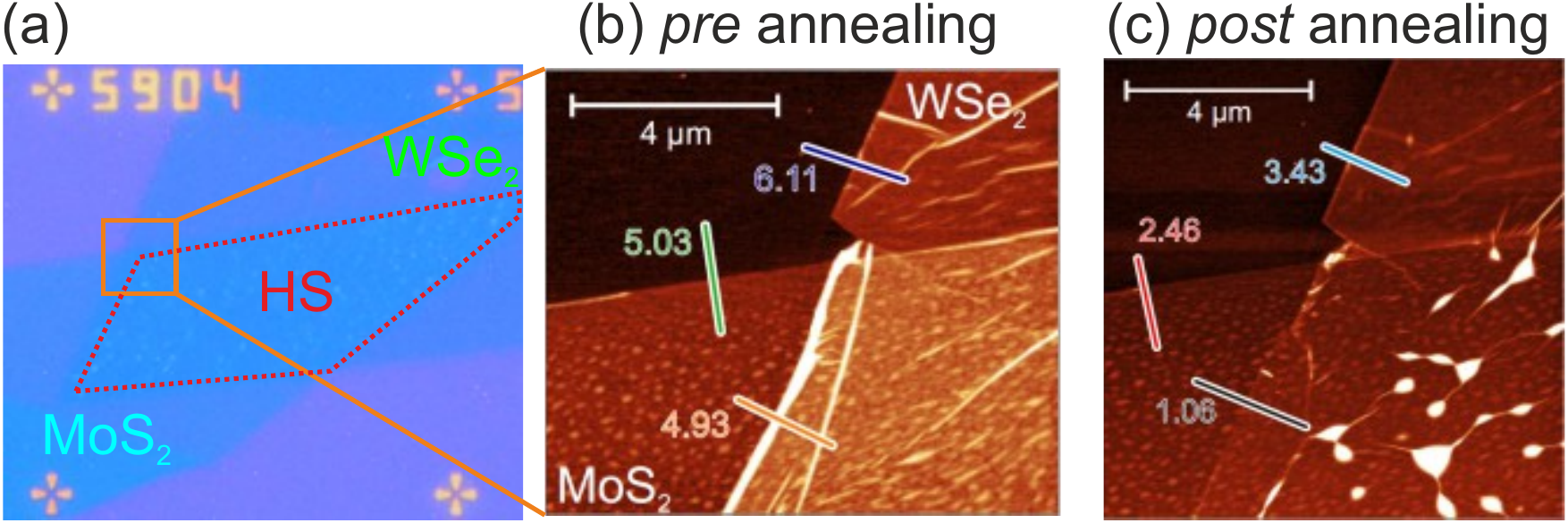}
\caption{\label{fig:Annealing_AFM}(a) Optical microscope image of heterobilayer sample. The orange rectangle marks the  area of the AFM scans shown in (b) and (c). (b) AFM scan of sample before annealing. (c) AFM scan of sample after annealing. The solid lines in (b) and (c) indicate line traces extracted from the AFM data, the corresponding numbers indicate step heights in nanometers.
}
\end{figure}

The effects of annealing are corroborated by AFM measurements taken on the same sample
before and after annealing. As Fig.~\ref{fig:Annealing_AFM}(b) demonstrates, the step heights from the substrate to the individual monolayers measured before annealing are on the order of a few nanometers, and the step height between the MoS$_2$ and  WSe$_2$ is of a similar magnitude. All of these values significantly exceed the thickness of a TMDC monolayer, which is less than one nanometer, indicating that adsorbates, such as hydrocarbons, are trapped beneath and in between the TMDC flakes and act as spacers, effectively decoupling the layers. After annealing, we find that all step heights are reduced (Fig.~\ref{fig:Annealing_AFM}(c)), and that the step height between the MoS$_2$ and  WSe$_2$ monolayers is close to the expected value of less than one nanometer. Additionally, we note the formation of various bubbles in the heterobilayer region, which indicates a coalescence of the trapped adsorbates~\cite{
Haigh2012_SI}.

\subsection{Vibronic interlayer coupling}
\begin{figure}
\centering
\includegraphics[width=0.5\textwidth]{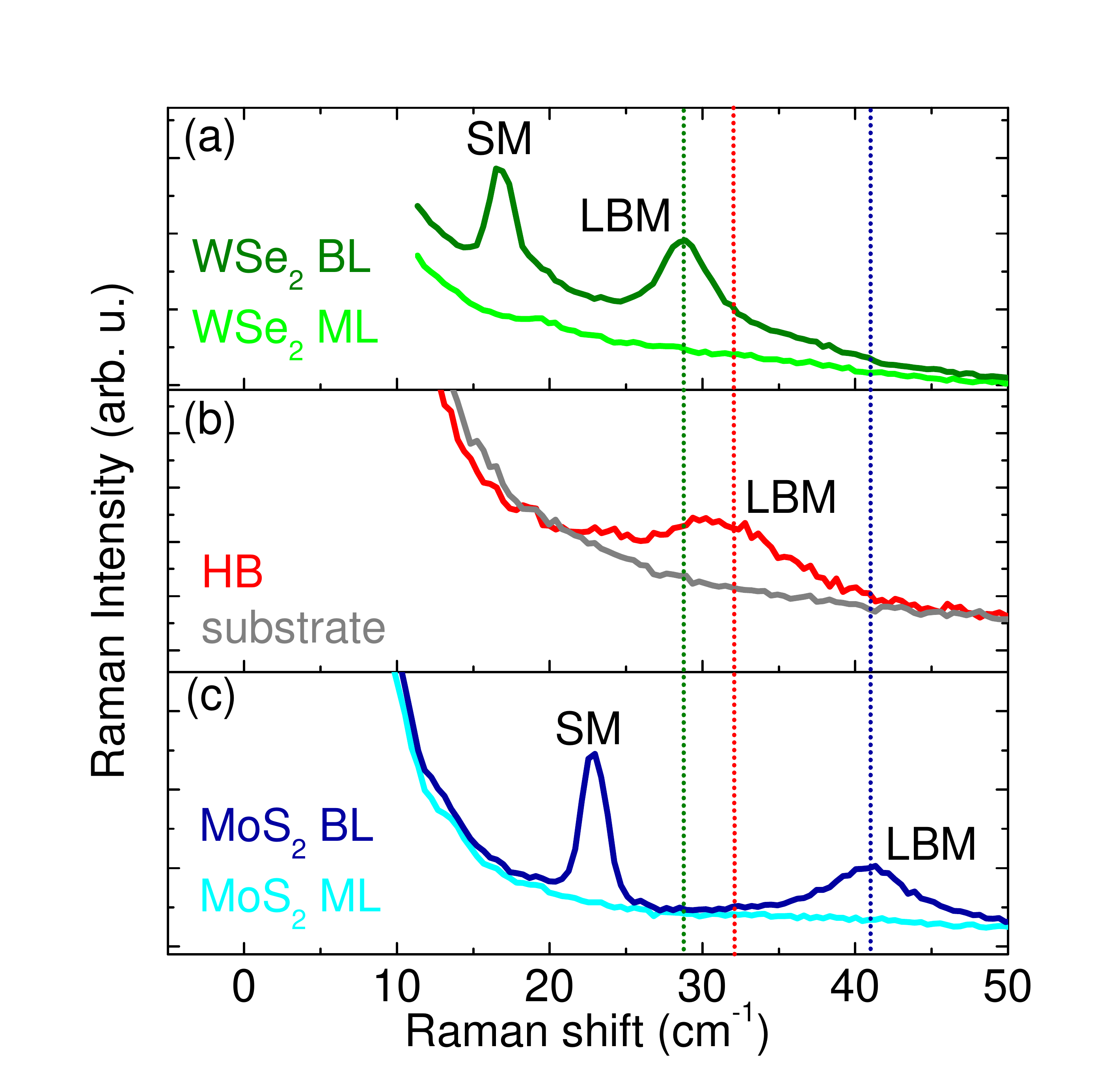}
\caption{\label{fig:Supp_Raman_LBM}{Low-energy} Raman spectra of (a) WSe$_2$ monolayers (ML)  and bilayers (BL), (b) heterobilayer (HB) and bare substrate, (c) MoS$_2$ mono- and bilayers. SM and LBM indicate the shear and layer breathing modes, respectively. The dotted lines mark the LBM positions in the three bilayers.
}
\end{figure}

Raman spectroscopy is a powerful tool to identify various two-dimensional crystals, and  to study the effects of external parameters such as strain and doping. While various two-dimensional materials have specific Raman modes dependent on the crystal structure, there are two modes that are generic to layered materials, the rigid-layer shear and breathing vibrations, where adjacent layers  oscillate with respect to each other. These rigid-layer oscillations
{have}
low Raman shifts due to the weak interlayer coupling. Figure~\ref{fig:Supp_Raman_LBM} shows a series of low-energy Raman spectra measured on WSe$_2$ (top)   and MoS$_2$ (bottom) mono- and naturally 2H-stacked bilayers, as well as on a heterobilayer. In the 2H-stacked bilayers, the  shear (SM) and breathing (LBM) modes are readily observable.  Naturally, both modes are absent in the monolayers. As expected from  its higher mass per unit area, SM and LBM modes in WSe$_2$ are at lower frequencies than in MoS$_2$. For both materials, nearly identical
spring constants $k$ have been extracted previously~\cite{Lui2014a_SI}. In the heterobilayer, we also observe the LBM mode, with a frequency  of 31~cm$^{-1}$, in between the values for WSe$_2$ (29~cm$^{-1}$) and MoS$_2$ (41~cm$^{-1}$) bilayers.
A  simple harmonic oscillator estimate assuming an averaged mass per unit area and an identical spring constant $k$ for the heterobilayer slightly overestimates the mode frequency, yielding a value of 35~cm$^{-1}$. By contrast, our DFT-TS calculations yield a value of 30~cm$^{-1}$, in excellent agreement with the measured value.
For these calculations the layer separation of twisted HB was slightly increased and decreased around the equilibrium value and the total energy of each geometry was calculated. The energy vs.~separation series was fitted to a harmonic function and the LBM frequency was calculated from the harmonic force constant.

\subsection{SHG microscopy}
\begin{figure}
\centering
\includegraphics[width=0.7\textwidth]{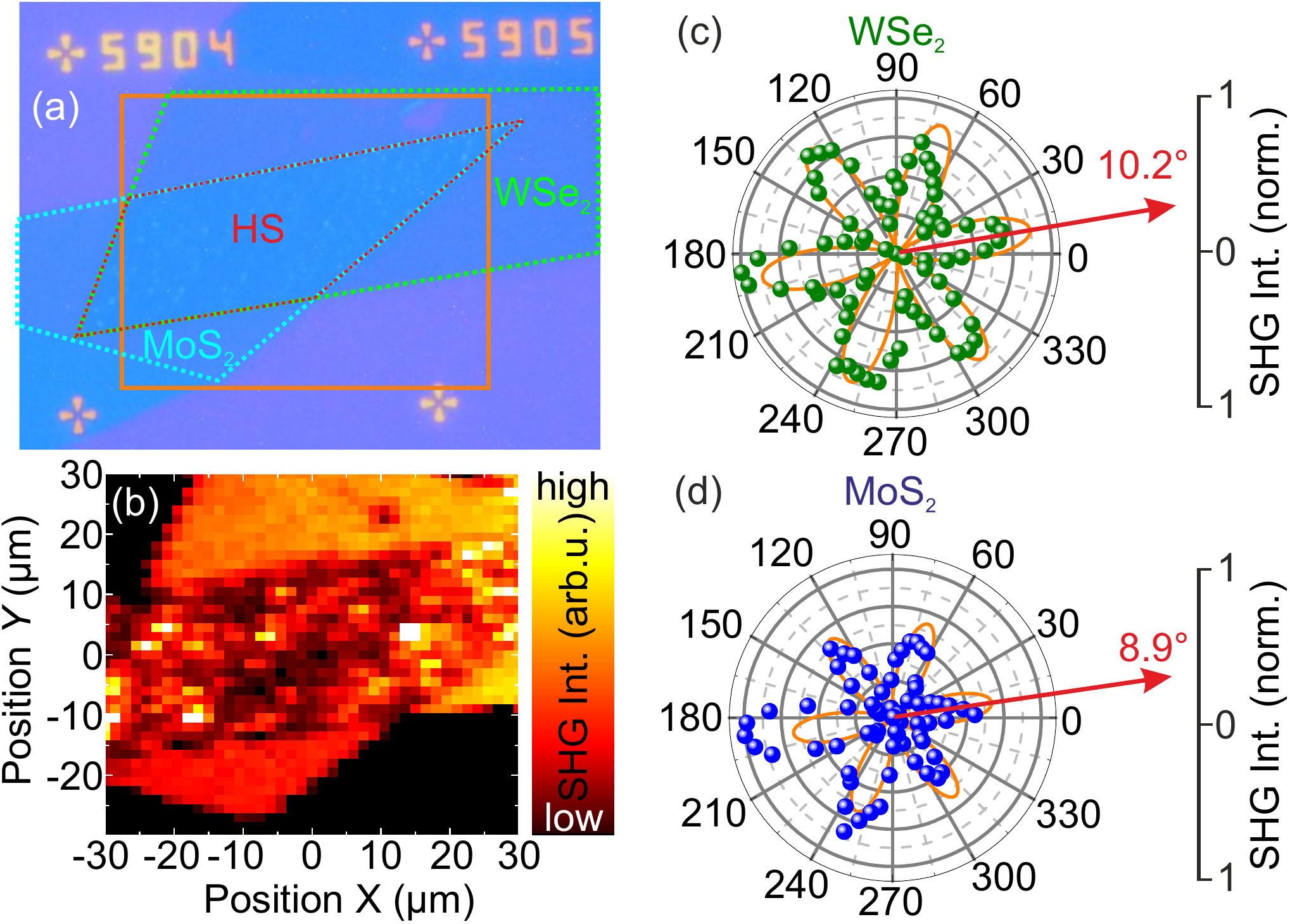}
\caption{\label{SHG_H1_4panel} (a) Optical microscope image of heterobilayer sample. The orange rectangle marks the  area of the SHG scan, shown in (b). The outlines of the MoS$_2$ and WSe$_2$ monolayers and the heterobilayer are marked by the dotted lines. (b) False color map of {total} SHG intensity as a function of position. (c) and (d) Polar plots of {the parallel component of the SHG intensity $I_{SHG}^{\|}$ as a function of  polarization} under linearly polarized excitation measured on the WSe$_2$ (c) and MoS$_2$ (d) monolayer regions. The red arrows indicate the orientation of an armchair direction relative to the horizontal polarization orientation, extracted from fits (solid orange lines) to the experimental data (dots). }
\end{figure}

In order to determine the twist angle of the heterobilayers, we employ polarization-resolved second-harmonic-generation (SHG) microscopy, which was previously shown to be a powerful tool to determine the crystallographic orientation of TMDC monolayers~\cite{Heinz_SHG13_SI,Malard_SHG13_SI} and stacking angles of heterobilayers~\cite{Hsu2014_SI}. Figure~\ref{SHG_H1_4panel} demonstrates this procedure for one particular sample, which is depicted in Fig.~\ref{SHG_H1_4panel}(a).  For each of the two constituent monolayers, we determine the relative angle $\phi_0$ between the armchair directions and the horizontal polarization axis of our SHG setup by measuring  the polarization-resolved SHG intensity $I_{SHG}$ on isolated monolayer regions. For parallel polarization of fundamental and SHG light, the SHG intensity is maximum if the polarization axis is oriented along an armchair direction, yielding a {$I_{SHG}^{\|}\propto \cos^2(3\phi-\phi_0)$} dependence. As Fig.~\ref{SHG_H1_4panel}(c) and (d) show, polar plots of
the SHG intensity can readily be fit to the expected {$I_{SHG}^{\|}$} dependence. However, the 6-fold symmetry of {$I_{SHG}^{\|}$} does not allow us to determine whether the monolayers in a heterobilayer are, e.g., aligned ($\alpha=0^{\circ}$) or antialigned ($\alpha=60^{\circ}$). To resolve this ambiguity, we perform scanning {measurements of the total SHG intensity}. In the heterobilayer region, SHG from the individual monolayers interfere constructively for aligned stacking and destructively for antialigned stacking. As shown in Fig.~\ref{SHG_H1_4panel}(b), we observe a reduced SHG intensity in the heterobilayer region, relative to the SHG intensities of the constituent monolayers, indicating destructive interference. We note that the SHG signal is not fully suppressed within the heterobilayer region due to the fact that the interfering SHG fields of the constituent monolayers have different amplitudes. {In conjunction with the polarization-dependent SHG measurements shown in Fig.~\ref{SHG_H1_4panel}(a),
 the destructive interference observed in the heterobilayer region allows us to determine a twist angle of $58.7^{\circ}$ for this sample.}

\subsection{Quenching of intralayer PL emission in heterobilayers}
\begin{figure}
\centering
\includegraphics[width=\textwidth]{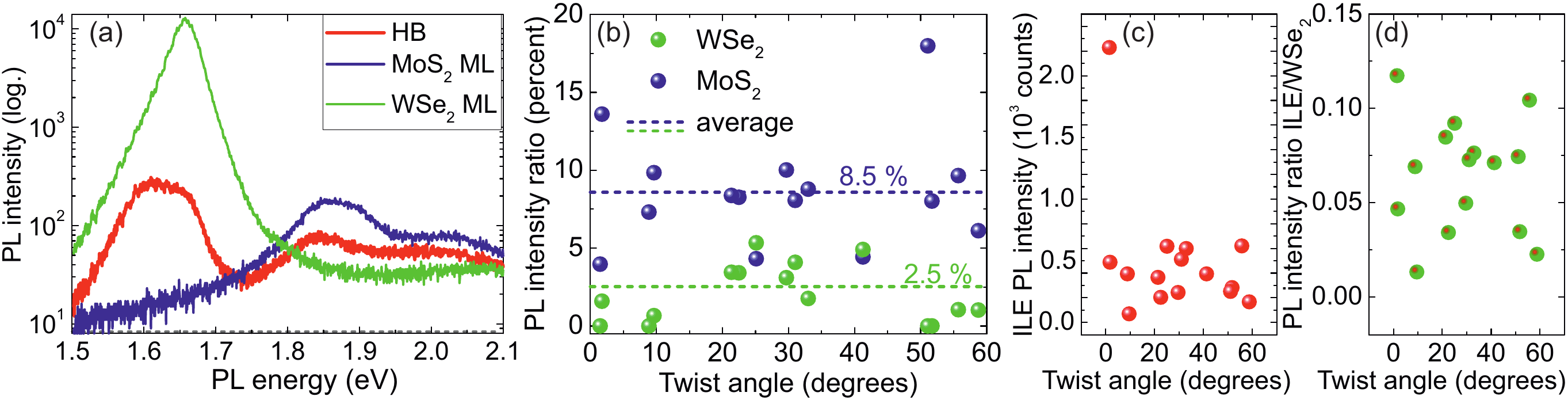}
\caption{\label{Quenching_VGL} (a) Room-temperature PL spectra measured on isolated  WSe$_2$  and MoS$_2$  monolayers and on a heterobilayer region. All spectra are recorded using identical excitation density and integration times. (b) Ratio of WSe$_2$ (green dots) and MoS$_2$ (blue dots) A {exciton/trion} PL intensities in the heterobilayer relative to isolated monolayers as a function of twist angle. The dashed lines indicate the ratio averaged over all twist angles. (c) ILE intensity as a function of twist angle. (d) Ratio of { ILE to WSe$_2$ A exciton intensity (measured on isolated monolayer region)} as a function of twist angle.
}
\end{figure}
In all of our heterobilayers, we observe a pronounced suppression/quenching of the monolayer A exciton PL emission of the constituent WSe$_2$  and MoS$_2$  monolayers, as shown in the main manuscript and in Fig.~\ref{Quenching_VGL}(a). Remarkably, we find that the quenching of the WSe$_2$ PL emission is far more pronounced than that of the MoS$_2$. This trend is observed for all of the investigated heterobilayer samples, as Fig.~\ref{Quenching_VGL}(b) demonstrates. While we do not find a clear dependence of this quenching factor as a function of twist angle, we see that  on average, the MoS$_2$ PL intensity is reduced to about 8~percent of the monolayer value, while the WSe$_2$ PL emission is reduced to about 2~percent. We note that due to the low intensity of the WSe$_2$ PL emission in the heterobilayer, combined with the spectral overlap with the ILE emission, there are larger fluctuations of the quenching ratio from sample to sample. While a quantitative analysis of the quenching ratios would require
measurements of interlayer tunneling rates from WSe$_2$  to MoS$_2$ and vice versa, as well as radiative recombination rates for the two materials, the large difference of the quenching ratios supports our calculations of the heterobilayer which yield a delocalized hole state that extends over the two constituent monolayers, and an electron state localized fully in the MoS$_2$ layer. In the heterobilayer, PL emission of MoS$_2$ A excitons may be enabled by \textbf{intralayer} hole scattering from the global valence band minimum at $\Gamma$ to the local minimum at $K$ and subsequent radiative recombination with electrons at the global conduction band minimum at $K$. By contrast, PL emission of WSe$_2$ A excitons from the heterobilayer would require  \textbf{interlayer} tunneling of electrons from the MoS$_2$ layer into the WSe$_2$ layer against a substantial band offset difference, leading to a more pronounced suppression of this transition in the heterobilayer.

\subsection{Twist angle dependence of interlayer exciton emission intensity}
Additional evidence for the $\Gamma$-$K$ character of the ILE emission {in WSe$_2$-MoS$_2$ heterobilayers} is the (lack of a) dependence of the ILE emission intensity on the twist angle.
In a twisted heterobilayer, both $\Gamma-K$ and $K-K$ transitions are generally indirect in $k$ space. However, for twist angles close to 0 and 60 degrees, the $K-K$ transition is almost direct in $k$ space, so that a larger PL yield could be expected for this transition. Figure~\ref{Quenching_VGL}(c) shows the ILE emission intensity as a function of twist angle. We note that the intensity fluctuates strongly from sample to sample and that there is no pronounced increase of PL yield for samples with twist angles close to 0 or 60 degrees. {This observation is in stark contrast to  WSe$_2$/MoSe$_2$ heterobilayers, where a clear dependency of the PL intensity of the ILE with respect to the twist angle was observed~\cite{Nayak2017_SI}}. In order to reduce sample-to-sample variations
which might stem from changes of the setup, such as laser spot size, we also determined the intensity ratio of the ILE emission {and the WSe$_2$ A exciton emission on the isolated monolayer regions}. As Fig.~\ref{Quenching_VGL}(d)
demonstrates, there is no clear dependence of this ratio on the twist angle either. This lack of an increased PL yield for near-aligned and antialigned heterobilayers supports our assertion that the ILE emission stems from the $\Gamma$-$K$ transition, which is, naturally, always indirect in $k$ space, irrespective of the twist angle.

\subsection{Energy shift of monolayer exciton/trion transitions in heterobilayers}
As discernible from {Fig.2(c)} of the main article, the A exciton/trion PL peaks do not exhibit a clear dependence on the twist angle. The PL energies from the ML parts are slightly scattered around well-defined mean values of 1.659 eV (WSe$_2$) and 1.890 eV, which are in good agreement with the literature \cite{Mak2010_SI,Zhao2013d_SI} and justify the identification as exciton/trion peaks.
However, we observe a redshift from the ML to the HB by 18 meV (WSe$_2$) and 24 meV (MoS$_2$). Similar redshifts in TMD homobilayers have been reported previously \cite{Mak2010_SI,Zhao2013d_SI}. It is likely to be a result of enhanced screening in the multilayers as compared to ML. The screening reduces both the quasiparticle band gap (redshift) and the A exciton binding energy (blueshift) and in total the redshift dominates~\cite{Latini2017_SI, Raja2017_SI}.

{
\subsection{Low-temperature PL measurements}
As discussed in the main manuscript, we observe a complex behavior of the ILE emission with decreasing sample temperature. Here, we present additional temperature-dependent PL measurements in the range from 90~K to 4~K. Figure~\ref{Fig_PL_temp}(a) shows a series of PL spectra measured on a heterobilayer and on isolated WSe$_2$ and MoS$_2$ monolayers. To suppress the effects of spatial inhomogeneity of the ILE emission, a PL map was recorded for each sample temperature, and spatial averaging of about 30 individual spectra measured in the heterobilayer region was performed. The spectra are not normalized, so that the PL intensities for different temperatures can be directly compared.
For both of the isolated monolayers, the systematic blueshift of the A exciton{/trion} PL emission with decreasing temperature due to band gap increase continues. We also observe that the A exciton and trion emission yield in the WSe$_2$ monolayer decreases as the temperature is lowered below 90~K. The origin of this decrease will be discussed below. Additionally, a broad shoulder (marked L) emerges at 60~K, which can be attributed to emission from localized states~\cite{Jones2013_SI,GangWang_SI}. This localized-state emission dominates the PL spectrum of the WSe$_2$ monolayer at 30~K and below.
In the MoS$_2$ monolayer, we find that the exciton/trion PL yield monotonously increases with decreasing temperature, most likely due to the suppression of nonradiative decay channels. We also observe the emergence of a broad, low-energy shoulder (marked S) as the temperature is reduced below 90~K. This feature was previously attributed to excitons bound to surface adsorbates~\cite{Plechinger2012_SI}.
In the heterobilayer region, we observe that, while the MoS$_2$ PL yield increases monotonously with decreasing temperature, the ILE emission is no longer clearly discernible at temperatures below 90~K. This is due to the emergence of the pronounced localized-state emission of WSe$_2$ in the heterobilayer region, which is at a slightly lower energy than the ILE, but so broad that it overlaps with the ILE emission region. Thus, we cannot clearly quantify the  ILE yield for temperatures below 90~K. However, an upper boundary for the remaining ILE emission can be extracted from the PL spectra. For this, we compare the ILE PL emission obtained at 90~K, blueshifted by the same amount as the MoS$_2$ emission to account for the changing transition energy, with the spectra at 60~K and below. Scaling the trace by an appropriate amount to match the PL spectra yields a conservative upper boundary for remaining ILE emission. This approach is indicated for the PL spectrum measured at 4~K by the red spectrum, which 
corresponds to the appropriately blueshifted ILE emission spectrum measured at 90~K, scaled by a factor of 0.7. We note that this upper boundary  most likely significantly exceeds the ILE emission that would be obtained by decomposition of the PL spectrum into Gaussian peaks for ILE and defect-related emission.
We also note that we do not observe the low-energy emission from MoS$_2$ in the heterobilayer region. This observation indicates that due to the {fact that} MoS$_2$ {is} covered by the WSe$_2$ layer, no pronounced adsorbate-related PL emission occurs.

\begin{figure}
\centering
\includegraphics[width=\textwidth]{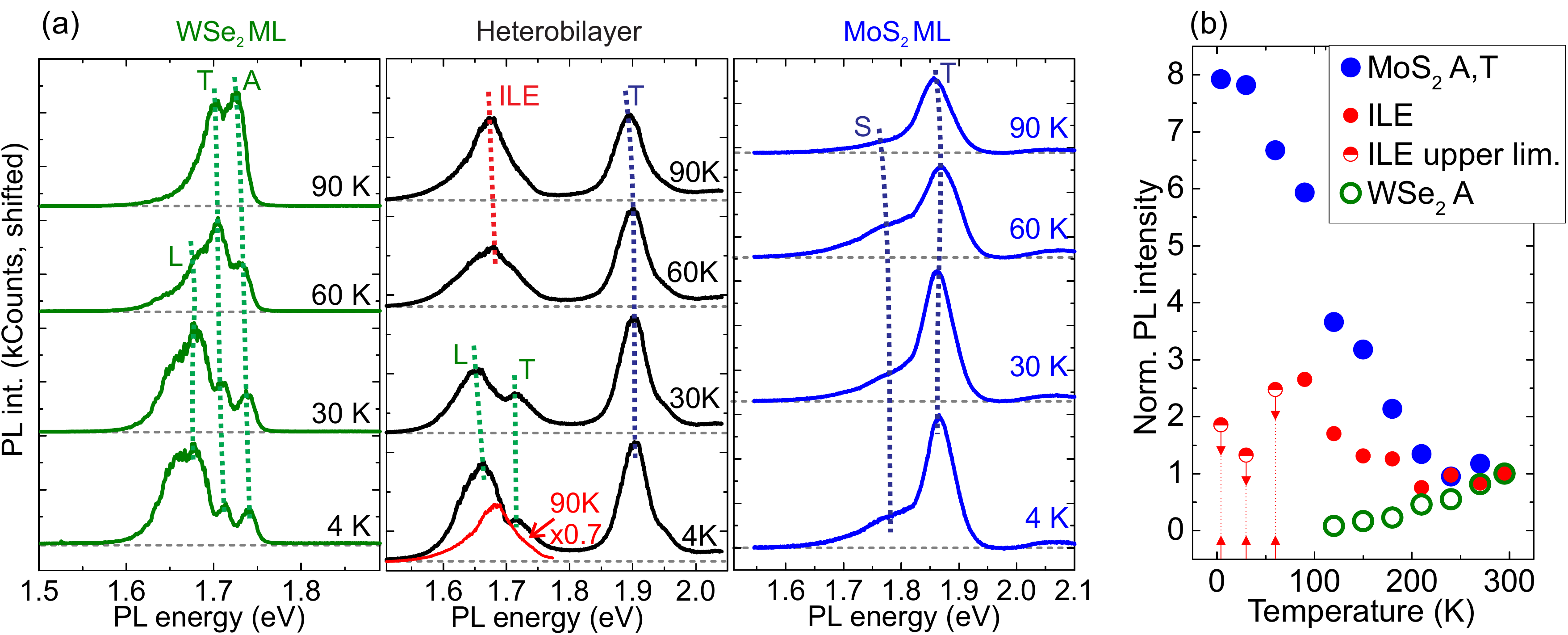}
\caption{\label{Fig_PL_temp} (a) Temperature-dependent PL spectra measured on isolated  WSe$_2$ (left panel)  and MoS$_2$ (right panel)  monolayers (ML) and on a heterobilayer region (center panel). {The spectra are not normalized, so that the PL intensities for different temperatures can be directly compared}. The dotted lines trace the spectral evolution of the various emission peaks. In the WSe$_2$ monolayer, emission from localized states (L) and trions (T) is observable in addition to the neutral A exciton. In the MoS$_2$ monolayer, a low-energy PL emission attributed to surface-adsorbate-bound excitons (S) is identified in addition to the {trion feature}. In the heterobilayer,the interlayer exciton (ILE) emission is suppressed as the temperature is reduced from 90~K to 60~K. As the temperature is reduced further, pronounced emission from the WSe$_2$ L peak masks any residual ILE emission. To estimate an upper boundary for residual ILE emission, the spectrum measured at 90~K is scaled by a factor of 
0.7 and superimposed onto the spectrum at 4~K, taking the blueshift in energy due to reduced temperature into account. (b) Integrated PL intensity of MoS$_2$ trion/exciton, WSe$_2$ exciton and ILE peaks, normalized to their respective intensities at 300~K, as a function of temperature. For the ILE emission intensities below 90~K, the estimated upper boundary is indicated.}
\end{figure}

In Fig.~\ref{Fig_PL_temp}(b), we plot the integrated PL intensities for the WSe$_2$ exciton, the MoS$_2$ exiton/trion and the ILE emission measured in the heterobilayer region, normalized to their respective values at 300~K, as a function of temperature.We find that the PL yield of the MoS$_2$-related emission monotonously increases with decreasing temperature. By contrast, for the WSe$_2$ A exciton emission,  the PL yield decreases with decreasing temperature, and the neutral A exciton emission is no longer observable at temperatures below 120~K. This decrease is due to the peculiar band structure of WSe$_2$, in which the conduction-band spin splitting is such that the optically bright, spin-allowed A exciton transition is between the upper valence and the upper conduction band. Hence, there is a lower-energy, optically dark A exciton state in WSe$_2$ and the related WS$_2$. The lower-energy dark A exciton state in the tungsten-based materials was indirectly inferred in temperature-resolved PL measurements, 
where the PL yield initially increases with increasing temperature due to increasing thermal population of the optically bright A exciton state~\cite{Zhang:2015d_SI,Arora:2015b_SI,Godde:2016a_SI}. More recently, PL emission from the dark state was directly observed using applied in-plane magnetic fields~\cite{Zhang:2017a_SI} and in-plane excitation and detection geometry~\cite{Wang:2017b_SI}. Since the ILE is formed from electrons located in the MoS$_2$ layer, where the conduction-band spin splitting is very small~\cite{Kormanyos2015_SI}, and holes that are located predominantly in WSe$_2$, we do not expect an energetically favorable, spin-forbidden (dark) ILE ground state, even if we were to assume a K-K nature of the ILE. Hence, we do not expect a nonmonotonous temperature dependence of the ILE emission based on dark state formation. However, we find that, after an initial increase of the ILE PL yield with decreasing temperature, most likely due to the suppression of nonradiative decay channels, we observe a clear 
suppression of the ILE emission as the temperature is decreased below 60~K. While we can only give an upper boundary for this suppression, this behavior is in stark contrast to that observed in the MoSe$_2$-WSe$_2$ heterobilayer system. For that material combination, in which ILE are considered to be k-space-direct,  a significant, monotonous increase of the ILE PL yield is observed with decreasing temperature~\cite{Rivera2014_SI,Nagler17_SI}. This clear difference in temperature dependence of the ILE PL yield further supports our assertion that the ILE observed in our MoS$_2$-WSe$_2$ heterobilayers are k-space-indirect. We note that we do not expect a complete suppression of the ILE PL emission in our system at low temperatures: while radiative recombination of k-space indirect excitons is partially suppressed at low temperatures due to the reduced phonon population, PL emission of k-space-indirect excitons remains possible via phonon emission processes.

The temperature-dependent PL data presented in the main manuscript and supplement were recorded on a sample with a twist angle of about 33~degrees. A second measurement series on a sample with near-alignment (twist angle about 59~degrees, not shown) yields a qualitatively identical behavior of the ILE PL intensity as a function of temperature, further supporting our assertion that the ILE transition in our heterobilayers is always k-space-indirect, irrespective of twist angle.
}

\newpage
\section{Supporting theoretical details}

\subsection{Model system}
\subsubsection{Basic electronic structure of heterobilayers}

\begin{figure}[tb]
\centering{\includegraphics[width=\textwidth]{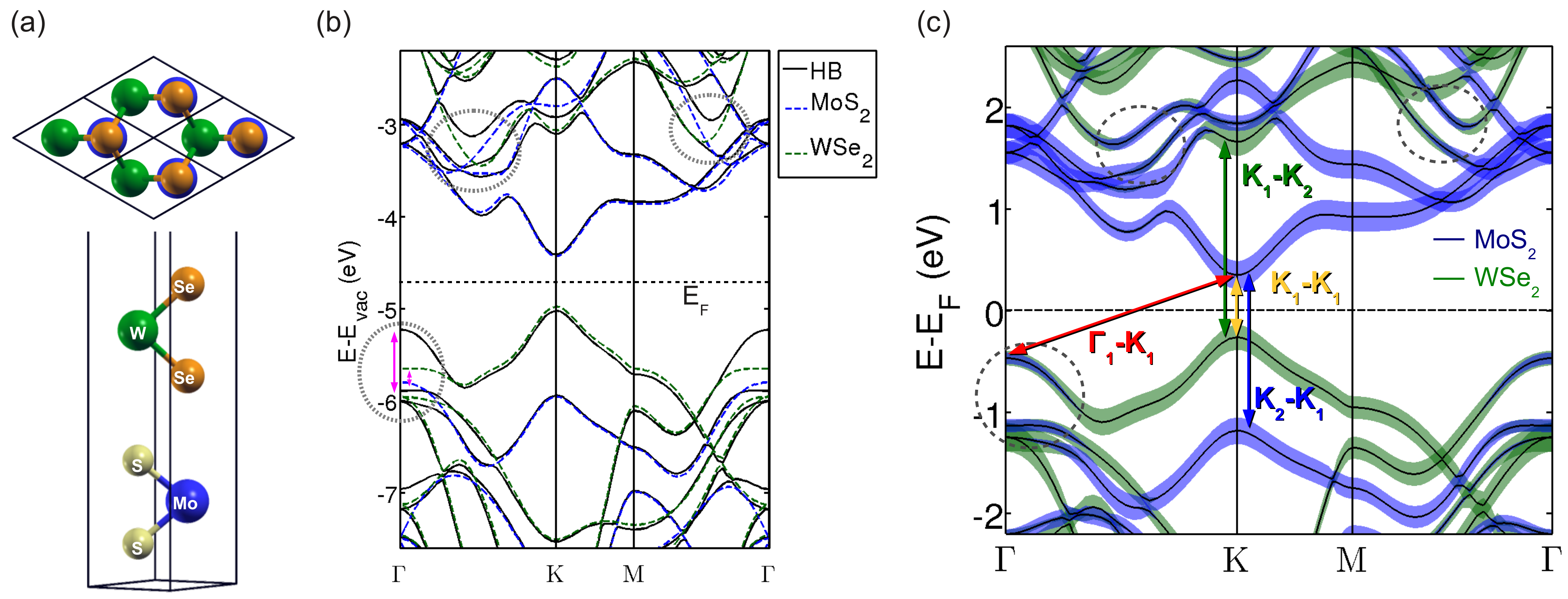}}
\caption{\label{fig:2H-basic}
(a) Top and side view of the strained, crystallographically antialigned and commensurate model system. Black lines show the hexagonal unit cell.
(b) Superposition of the band structures of the HB model system (solid, black) and the individual (strained) layers of MoS$_2$ (dashed, blue) and WSe$_2$ (dashed, green). Grey circles indicate regions where interlayer hybridization leads to strong modifications of the band structure.
(c) Projected band structure of the HB model. The width of the bands is proportional to the monolayer character of the state. Spectroscopically relevant transitions are indicated by arrows; states are labeled as in Fig.~\ref{fig:2H-sep}(a). Encircled regions again indicate interlayer hybridization.
}
\end{figure}

In order to understand the basic properties of WSe$_2$/MoS$_2$ HBs we first analyze an idealized  model system, where the MLs are crystallographically antialigned and commensurate.
In this system the crystallographic antialignment implies a twist angle of 60\degree as in TMD bulk crystals of the 2H phase.
Lattice commensurability is enforced by an in-plane compression of 1.6\% for the WSe$_2$ ML and  an expansion of 2.1\% for MoS$_2$ (the in-plane DFT+TS lattice constants of MoS$_2$ and WSe$_2$ are 3.16 {\AA} and 3.28 {\AA}, respectively). The model system is illustrated in Fig.~\ref{fig:2H-basic}(a).
Its electronic band structure is shown in  Fig.~\ref{fig:2H-basic}(b), where the band structures of the individual MoS$_2$ and WSe$_2$ layers are superimposed and all energy levels are aligned with respect to the vacuum level $E_\mathrm{vac}$. The latter was determined from the constant value of the DFT effective potential (without the exchange-correlation potential) at large distances from the system.
Due to the strong applied strain, band gaps, band extrema and band dispersions are  modified in comparison to {strain}-free systems \cite{Peelaers2012_SI,Ghorbani-Asl2013_SI}. For example,
the valence band maximum of ML MoS$_2$ in Fig.~\ref{fig:2H-basic}(b) is at the $\Gamma$-point. However, in strain-free ML MoS$_2$ it is at the K point and the band gap is also bigger (and that of WSe$_2$ is smaller).
We will use the model system to analyze the influence of interlayer interactions on the band structure.
As discernible in Fig.~\ref{fig:2H-basic}(b) the band structure of the HB is mostly a superposition of the band structures of the individual MLs. The staggered band alignment is reflected in the valence band maximum being defined by WSe$_2$ states and the conduction band minimum by MoS$_2$ states. The localization of the states is indicated in Fig.~\ref{fig:2H-basic}(c) and it is apparent that wherever the band energies agree with the corresponding ML value in Fig.~\ref{fig:2H-basic}(b) the states are fully localized in {the respective} layer.
However, electronic interlayer interactions modify certain states of the HB band structure. Their band energies are shifted and hybrid interlayer states are formed. These states are indicated by dashed, grey circles in Fig.~\ref{fig:2H-basic}(b) and (c). Most relevant for this study is the interlayer hybridization at the valence band maximum near the $\Gamma$-point. Pink arrows indicate the strong shift of band energies. Important is the upshift of the highest occupied band. The band curvature increases, which implies a reduction of the hole effective mass. The wave function of this state is a hybrid of 76\% WSe$_2$ and 24\% MoS$_2$.
The second{-}highest occupied band at $\Gamma$ is shifted down and its band curvature decreases (increase of the hole effective mass).

\subsubsection{Identification of the nature of ILE via the dependence of the electronic structure on the layer separation}

\begin{figure}[tb]
(a) \includegraphics[width=.50\textwidth,trim=0 -10 0 0,clip]{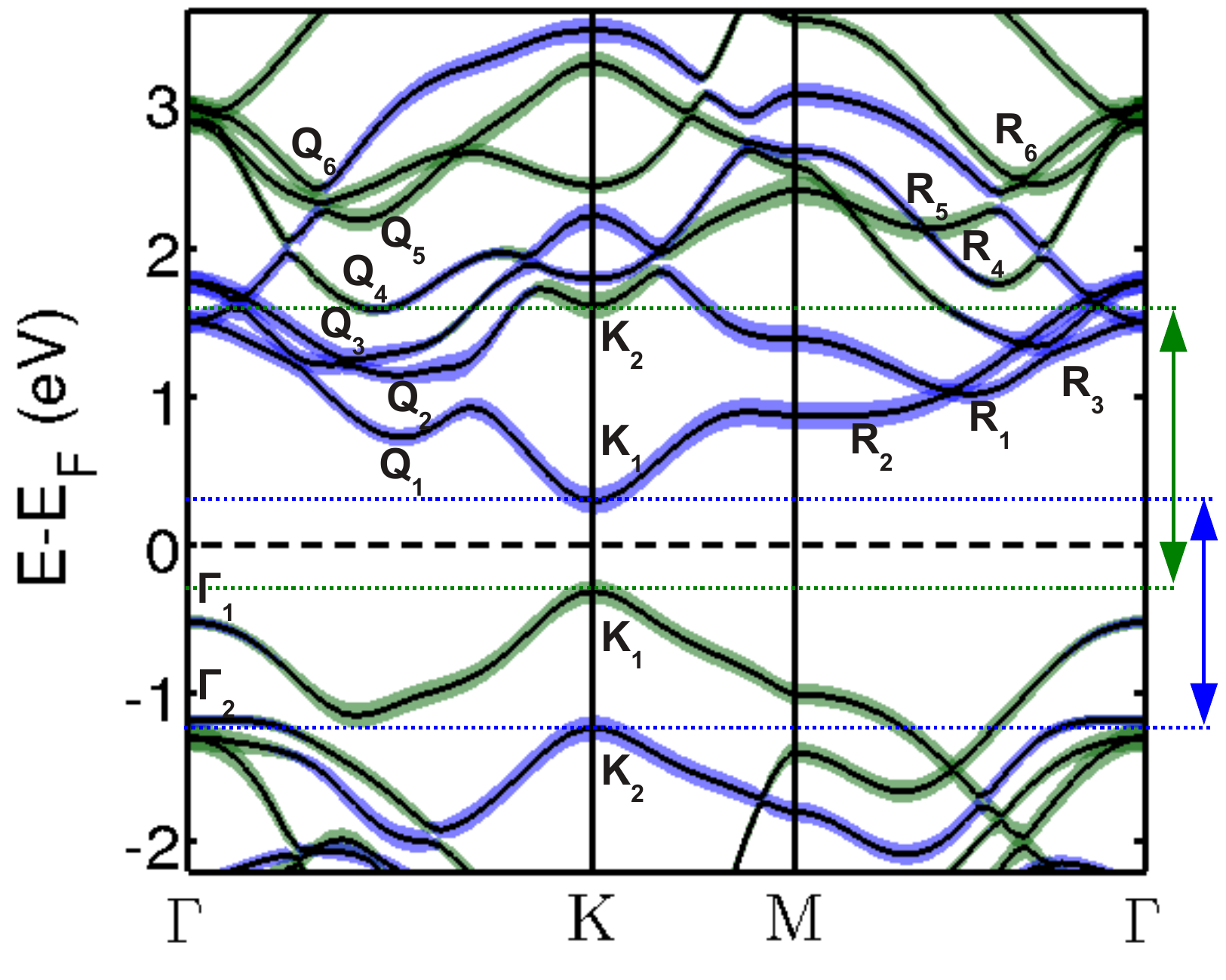}
(b) \includegraphics[width=.45\textwidth]{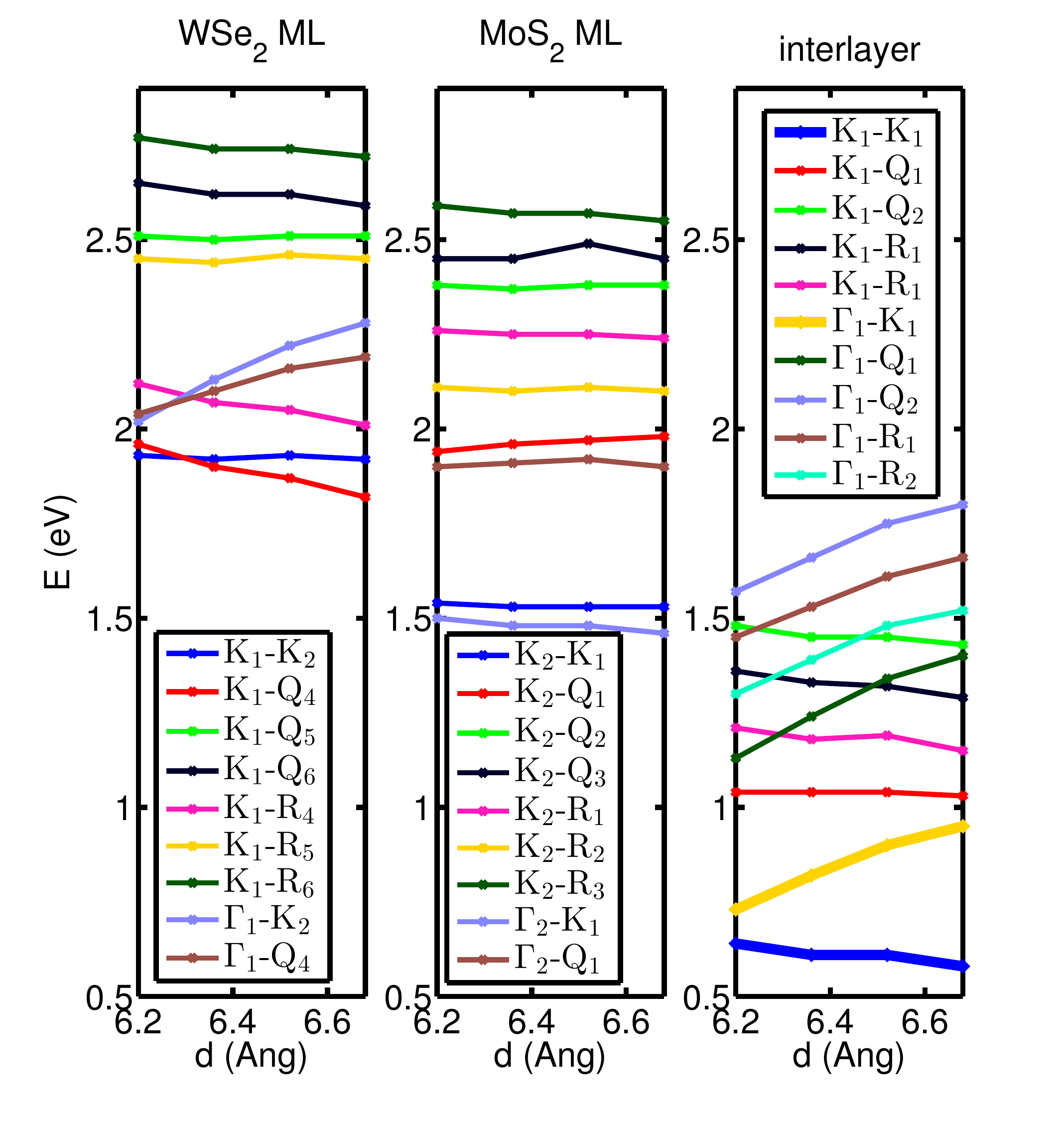}
\caption{\label{fig:2H-sep}
The dependence of interlayer transition energies on the layer separation in the model system.
(a) The labeling of band extrema near the valence and conduction band edges. Vertical arrows indicate the monolayer band gaps. Color code as in Fig.~\ref{fig:2H-basic}(c).
(b) DFT transition energies as function of interlayer separation $d$ for monolayer (ML) and interlayer transitions. The transitions are defined as $k^\mathrm{valence}_i-k^\mathrm{conduction}_j$.
}
\end{figure}

With PL spectroscopy we found that the PL peak near 1.6 eV (dubbed `ILE') shifts in energy as a continuous function of the twist angle. Furthermore a DFT analysis of the HB
has shown that the separation between the layers changes as function of the twist angle. Most remarkably, the functional dependence of both effects on the twist angle is very similar with a linear proportionality factor of 0.44 eV/{\AA} (see main article).
In order to explain these PL results we need to identify a low-energy band transition that increases in energy as the layer separation increases (i.e. positive slope).
Therefore we study the dependence of the DFT transition energies on the separation between the ML. The results are given in Fig.~\ref{fig:2H-sep}. Panel (a) is similar to Fig.~\ref{fig:2H-basic}(c) but additionally the band extrema near the gap are labeled. Transitions between states of the valence and the conduction band in panel (b) are defined as $k^\mathrm{valence}_i-k^\mathrm{conduction}_j$.
We emphasize again that the ML of the model system are modified by in-plane strain and therefore band gaps and absolute transition energies are different from  strain-free systems (see above). However to a first approximation the out-of-plane interlayer interactions can be considered as  independent of the in-{plane} strain. Therefore, general trends  of the effect of layer separation on the electronic structure can be studied with the model system.

Within the considered range all transition energies shift as a nearly linear function of the layer separation and the extracted slopes are given in Table \ref{tab:slopes}.
ML transitions (left and central panel in Fig.~\ref{fig:2H-sep}(b)) that involve hybridized states ($\Gamma_1$, Q$_4$, R$_4$) exhibit a pronounced dependence on the layer separation, while transitions between non-hybridized states (K$_i$, Q$_1$, Q$_2$, Q$_3$, Q$_5$, R$_1$, R$_2$, R$_3$, R$_5$) do not. This is plausible since the interlayer hybridization is tuned by the layer separation.
MoS$_2$ ML transitions involve mostly non-hybridized states and therefore they are almost insensitive to the layer separation (the slopes are nearly zero).
WSe$_2$ ML transitions that involve $\Gamma_1$, Q$_4$, R$_4$ exhibit a separation dependence. But only $\Gamma_1$-related ones ($\Gamma_1$-K$_2$ and $\Gamma_1$-Q$_4$) have a positive slope (0.561 and 0.319 eV/{\AA}, respectively). However the transition energy is larger than that of the K$_1$-K$_2$ ML transition (also in unstrained systems), which is related to the A exciton.
Therefore these transitions cannot be measured with PL and we exclude the possibility that the PL peak near 1.6 eV is related to  transitions within ML of MoS$_2$ or WSe$_2$.
As a result of the staggered band alignment, all interlayer transitions (right panel in Fig.~\ref{fig:2H-sep}(b)) are at smaller energies than ML transitions. The PL peak near 1.6~eV is therefore likely to be related to an interlayer transition.
Transitions with positive slopes (last line in Tab.~\ref{tab:slopes}) are only those that involve $\Gamma_1$. The slopes of $\Gamma_1$-K$_1$, $\Gamma_1$-Q$_2$, $\Gamma_1$-R$_1$, $\Gamma_1$-R$_2$ are all very close to 0.44 eV/{\AA}. But $\Gamma_1$-K$_1$ is the lowest{-}energy transition and the only one of them that is relevant for PL measurements. We emphasize that also in {the} unstrained system K$_1$ forms the conduction band edge and Q$_2$, R$_1$ and R$_2$ are at higher energies.
Thus, we are able to identify a single transition -- $\Gamma_1$-K$_1$ -- that lies within a reasonable energy range, has the right trend and the correct positive slope of 0.47~eV/{\AA}, in excellent agreement with the value of 0.44~eV/{\AA}, found for realistic HB.

It is interesting to note the negative slope of the $K_1-K_1$ interlayer transition in Tab.~\ref{tab:slopes}. This means that the $K_1-K_1$ gap of isolated ML is smaller than the one in the HB. A similar $K_1-K_1$ gap reduction between WSe$_2$/MoS$_2$ HBs and their isolated ML was reported by Latini et al.~\cite{Latini2017_SI}. They ascribe this effect to charge transfer between the ML.

\begin{table}[tb]
\begin{tabular}{l|llllllllllllllllllllllllllllllllllll}
WSe$_2$ ML & K$_1$-K$_2$ & K$_1$-Q$_4$ & K$_1$-Q$_5$ & K$_1$-Q$_6$ & K$_1$-R$_4$ & K$_1$-R$_5$ & K$_1$-R$_6$ & $\Gamma_1$-K$_2$ & $\Gamma_1$-Q$_4$ \\
(eV/{\AA})   & -0.025 & -0.267 & 0.013 & -0.112 & -0.223 & 0.000  & -0.097 & 0.561 & 0.319 \\
\hline
MoS$_2$ ML & K$_2$-K$_1$ & K$_2$-Q$_1$ & K$_2$-Q$_2$ & K$_2$-Q$_3$ & K$_2$-R$_1$ & K$_2$-R$_2$ & K$_2$-R$_3$ & $\Gamma_2$-K$_1$ & $\Gamma_2$-Q$_1$ \\
(eV/{\AA})   &-0.019 & 0.071 & 0.002 & 0.023 & -0.037 & -0.017 & -0.076 & -0.081 & 0.010 \\
\hline
interlayer & \textbf{K$_1$-K$_1$} & K$_1$-Q$_1$ & K$_1$-Q$_2$ & K$_1$-R$_1$ & K$_1$-R$_2$ & {\textbf{$\boldsymbol{\Gamma}_1$-K$_1$}} & $\Gamma_1$-Q$_1$ & $\Gamma_1$-Q$_2$ & $\Gamma_1$-R$_1$ & $\Gamma_1$-R$_2$\\
(eV/{\AA}) & \textbf{-0.116} & -0.025 & -0.094 & -0.133 & -0.113 & \textbf{0.471} & 0.561 & 0.492 & 0.453 & 0.474\\
\end{tabular}
\caption{The slopes extracted from the transition energy vs. layer separation plots in Fig.~\ref{fig:2H-sep}(b).
}
\label{tab:slopes}
\end{table}	
	
\subsection{Twisted heterobilayers}
\subsubsection{Construction of twisted heterobilayers}

\begin{figure}[tb]
\includegraphics[width=\textwidth]{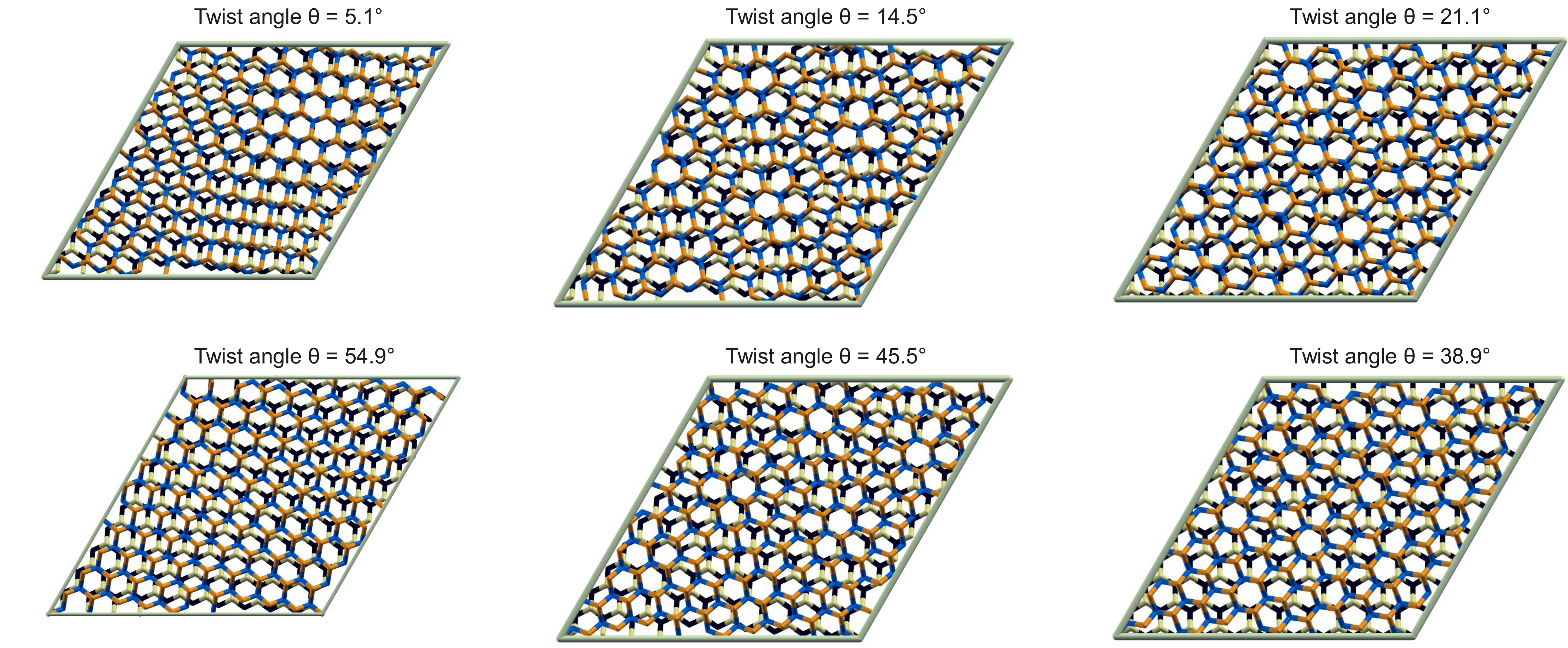}
\caption{\label{fig:HBCells}
Unit cells (green) and atomic decoration (Mo is black, S is yellow, W is blue and Se is orange) of the heterobilayers studied with DFT. MoS$_2$ is the bottom layer and WS$_2$ is the top layer. The wavelength of the discernible moir\'{e} patterns increases from the right column (twist angles close to 30\degree) to the left column (twist angles close to 0\degree or 60\degree).
}
\end{figure}

In order to generate the atomic structure of twisted HBs, commensurate, periodic supercells for special twist angles were constructed. The unit cells of the considered systems are shown in Fig.~\ref{fig:HBCells}.
The supercell lattice vectors for the top and bottom layers are given by
\begin{align*}
 \vec{T} = n_t \ \vec{t}_{1} + m_t \ \vec{t}_{2} \\
 \vec{B} = n_b \ \vec{b}_{1} + m_b \ \vec{b}_{2},
\end{align*}
where $\vec{t}_{1}$ and $\vec{t}_{2}$ are the primitive lattice vectors for the top layer, $\vec{b}_{1}$ and $\vec{b}_{2}$ are the primitive lattice vectors for the bottom layer (hexagonal lattice vectors) and $n_t$, $m_t$, $n_b$ $m_b$ are {integers}. The commensurability condition then simply is $|\vec{T}| = |\vec{B}|$.
As the DFT-TS lattice constants of MoS$_2$ and WSe$_2$ (3.16 and 3.28~{\AA} \cite{Opalovskii1965_SI,Agarwal1979_SI}) differ by about 4\%, the systems are actually incommensurate. However various supercells of one layer can be constructed such that they are nearly commensurate to supercells of the other layer. Commensurability is then enforced by applying a small amount of strain to the individual layers.
In our study the strain is always less  than $0.03 \%$ (see Table \ref{tab:HBCells}). DFT calculations of positive biaxial strain in MoS$_2$ show that the band gap changes by -0.240 eV/\% strain. The strain in the commensurate HB supercells therefore alters the MoS$_2$ band gap by less than 10 meV. This is much smaller than the band structure effects we are considering in this work.
All relevant structural parameters are given in Table \ref{tab:HBCells}. Each set of parameters describes a small and a large angle HB. The differences is generated by a similar or opposite relative orientation of the transition and chalcogen atoms in the primitive unit cells of each layer.

\begin{table}[tb]
\begin{tabular}{|l|l|r|l|r|r|r|}
\hline
 & \multicolumn{2}{c|}{WSe$_2$} & \multicolumn{2}{c|}{MoS$_2$} \\ \hline
twist angle (\degree) & ($n_t$, $m_t$) & strain (\%) & ($n_b$, $m_b$) & strain (\%) & atoms & $|\vec{T}|=|\vec{B}|$ (\AA) \\ \hline
5.1, 54.9 & (9,2) & 0.029 & (10, 1) & -0.029 & 642 & 33.277 \\ \hline
14.5, 45.5 & (2,9) & 0.029 & (10, 1) & -0.029 & 642 & 33.277 \\ \hline
21.1, 38.9 & (7,4) & -0.027 & (10, 0) & 0.027 & 579 & 31.602 \\ \hline
\end{tabular}
\caption{The parameters defining the commensurate supercells of the HB for each twist angle. The primitive-to-supercell scaling factors $n_t$, $m_t$, $n_b$ and $m_b$, the strain in each layer, the number of atoms per HB supercell and  the length of the supercell basis vector are given.}
\label{tab:HBCells}
\end{table}

\subsubsection{\label{sec:separation}Layer separation and static waviness}
\begin{figure}[bt]
(a)\includegraphics[width=.25\textwidth]{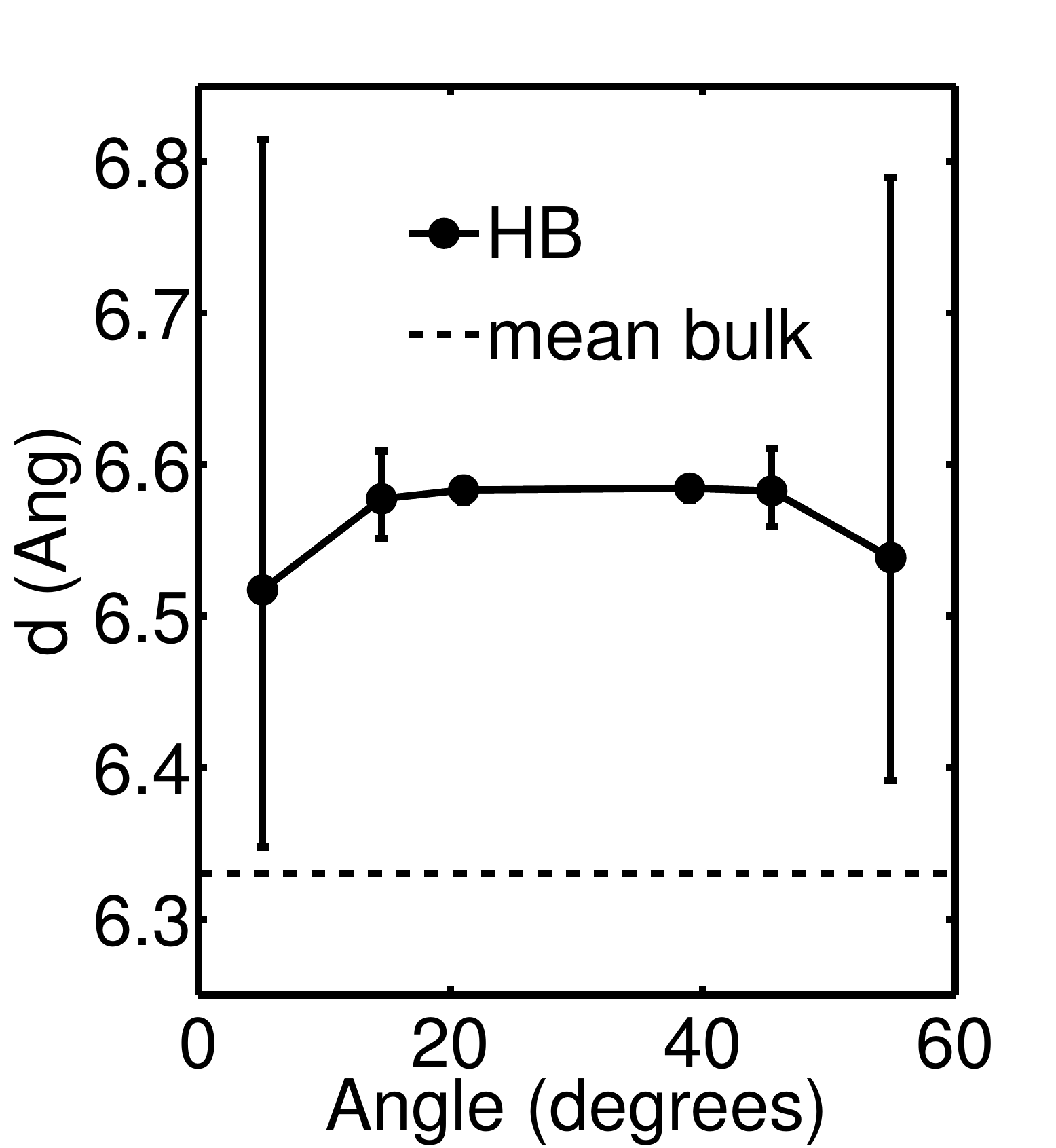}
(b)\includegraphics[width=.7\textwidth]{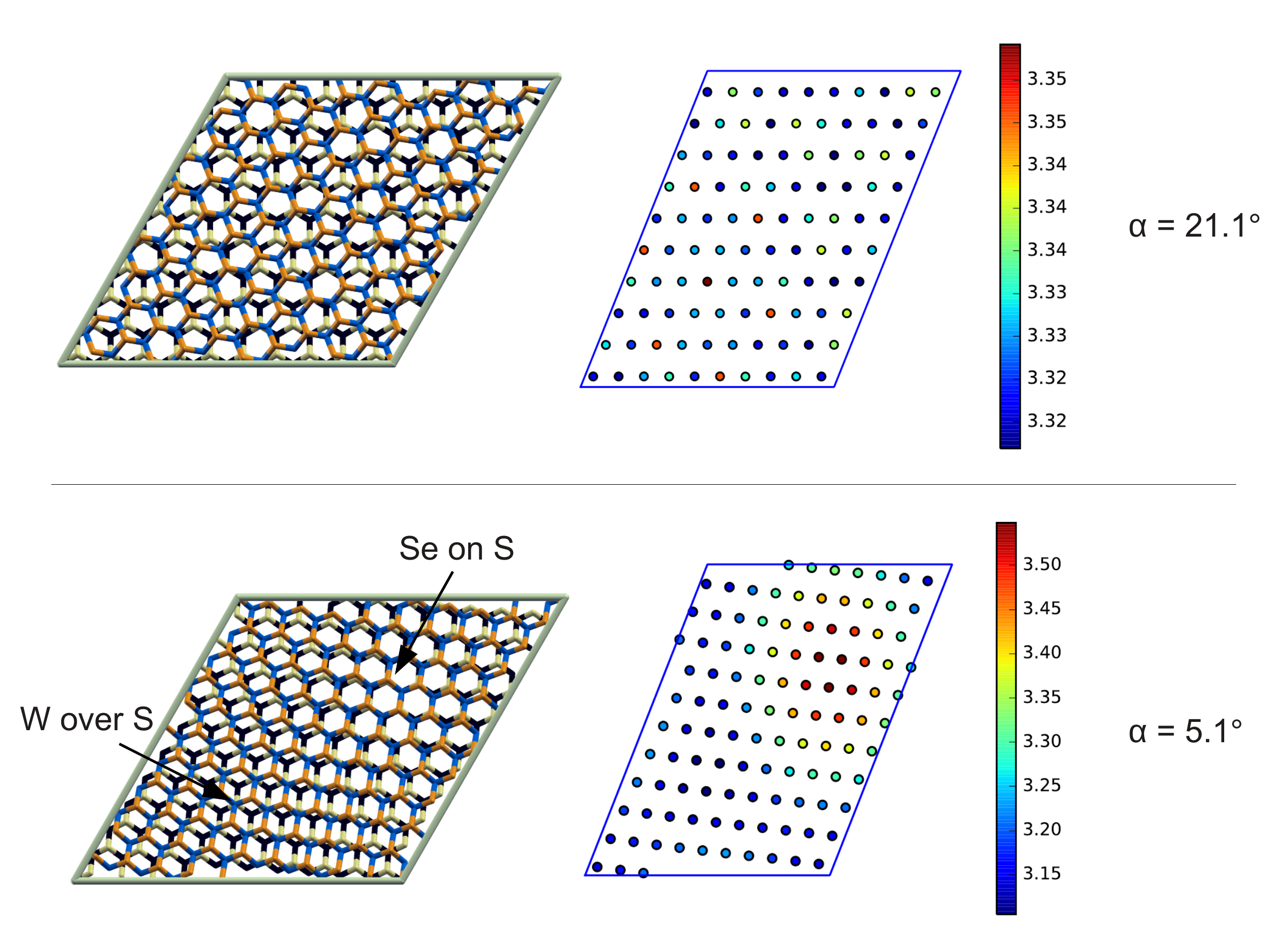}
\caption{\label{fig:waviness}
(a) Mean layer separation of MoS$_2$-WSe$_2$ heterobilayers vs.~twist angle. The `error bars' indicate the minimum and maximum local variation of the separation that becomes very strong for angles near 0\degree and 60\degree.
(b) left: structure illustration (color code as in Fig.~\ref{fig:HBCells}); right: color map of S-Se distances (vertical components). Distances in the color bar are in {\AA}.
}
\end{figure}

The twisted HB were structurally optimized with dispersion-corrected DFT-TS calculations.
Figure \ref{fig:waviness}(a) shows the mean layer separation $d$ of the considered HB. The latter changes as a continuous function of the twist angle over a range of 0.07 {\AA}.
Incommensurability and steric effects lead to an 4\% increase of $d$ as compared to a hypothetical commensurate HB system (dashed line in Fig.~\ref{fig:waviness}(a) at $d=6.33$ {\AA}, taken as half of the mean lattice constants of 2H MoS$_2$ and WSe$_2$).
The reduction of $d$ for angles near 0\degree or 60\degree by 1\% is a result of static waviness.
This effect is illustrated in Fig.~\ref{fig:waviness}(b).
The upper panel shows a HB with a twist angle of 21.1\degree and the upper right panel displays the vertical components of S-Se distances and how they vary within the unit cell. One finds  variations in a range between 3.32 and 3.35 {\AA}, which is  insignificant.
This is very different for the HB with twist angles near 0\degree or 60\degree.
In the 5.1\degree HB in the lower panel a long-wavelength \moire pattern is discernible in the structure illustration on the left. The local layer separation is maximal if two chalcogen atoms are vertically aligned (arrow 'Se on S') and minimal if chalcogen and metal atoms are vertically aligned (arrow 'W over S'). This is the essence of steric effects in TMDC {homobilayers and} HB. The S-Se distances in the lower right panel show a large variation, ranging from 3.10 {\AA} to 3.55 {\AA}. Similar elastic deformations in van der Waals HS have been reported by {some of the authors}~\cite{VanderZande2014_SI} and Kumar et al.~\cite{Kumar2015a_SI}.
Due to the long-wavelength of the \moire pattern and a flexibility of the ML to bend, these two extrema are realized within a single unit cell and create a static waviness.
For 21.1\degree the wavelength of the \moire pattern of the HB is too small and the two ML are too stiff to conform it and therefore the HB is not wavy and the S-Se distances are close to {3.34~{\AA}}.
For 5.1\degree the S-Se distances are smaller than 3.34 {\AA} in large parts of the unit cell. This is the reason why the mean layer separation is reduced for small twist angles.

\subsubsection{Determination of DFT transition energies}
\begin{figure}[bt]
(a)\includegraphics[width=.55\textwidth]{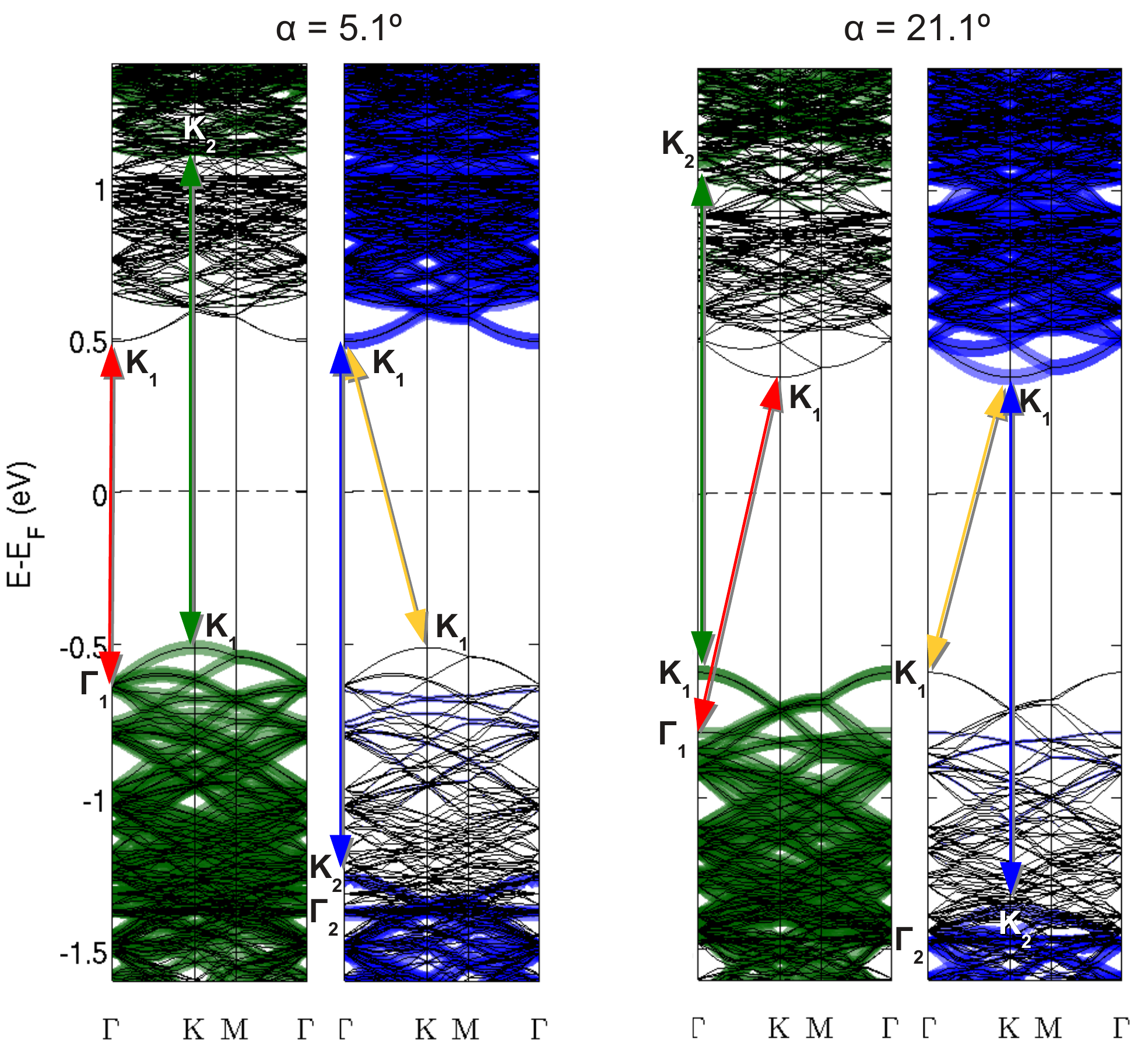}
(b)\includegraphics[width=.4\textwidth]{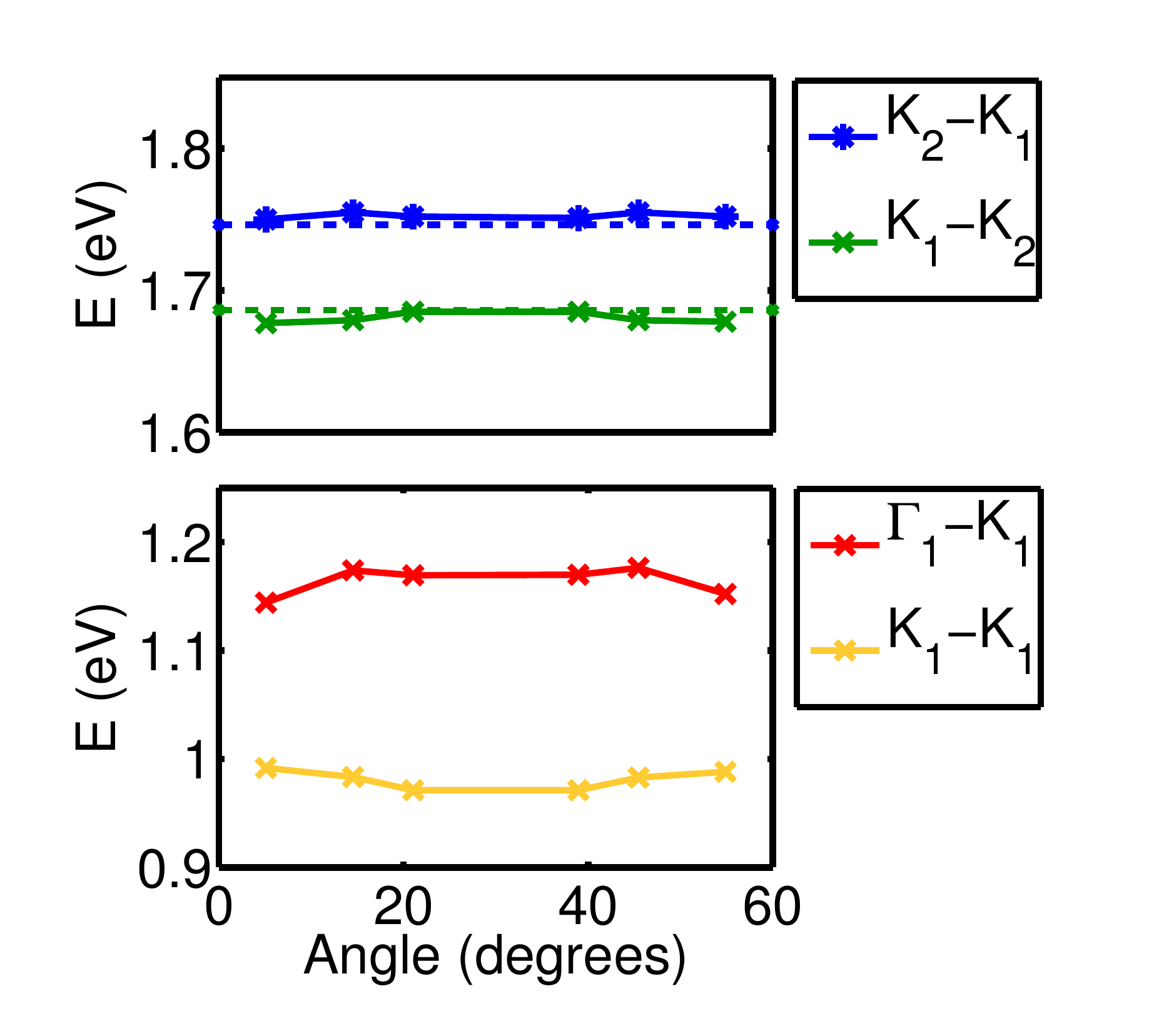}
\caption{\label{fig:transitions}
(a) DFT band structures of HBs with twist angles of 5.1\degree (left) and 21.1\degree (right). The color indicates the localization of the band states in the WSe$_2$ layer (green) and the MoS$_2$ layer (blue); each band structure is doubled for clarity. Relevant optical transitions are indicated by colored arrows. States are labeled as in Fig.~\ref{fig:2H-sep}(a).
(b) Important DFT transition energies for HB with varying twist angles. Dashed lines in the upper panel indicate values obtained from isolated monolayers.
}
\end{figure}

The band structures of two twisted HB are shown in Fig.~\ref{fig:transitions}.
Due to the large number of atoms per cell (ca. 600) the number of bands is also very large and the complexity does not allow an easy analysis as done with the HB model system (compare Fig.~\ref{fig:2H-basic}).
In twisted HB the Brillouin zones of the individual ML are twisted with respect to each other (see main article) and their bands are folded into the Brillouin zone of the commensurate supercell. This folding also implies that states from the $K$-points of the ML are folded to  general $k$-points in the HB supercell.
By tracing the mapping of the ML $K$ points, we find for the two examples, i.e. 5.1\degree HB:  K$_\mathrm{MoS_2} \rightarrow \Gamma_\mathrm{HB}$, K$_\mathrm{WSe_2} \rightarrow$ K$_\mathrm{HB}$; 21.1\degree HB: K$_\mathrm{MoS_2} \rightarrow$ K$_\mathrm{HB}$, K$_\mathrm{WSe_2} \rightarrow \Gamma_\mathrm{HB}$.
Note that the $\Gamma$-point always maps to $\Gamma$: $\Gamma_\mathrm{ML} \rightarrow \Gamma_\mathrm{HB}$.
It is straightforward to extract the energy of the $K_1-K_1$ transition from the valence band maximum (VBM) and conduction band minimum (CBM) in Fig.~\ref{fig:transitions}(a).
To find the energy of the VBM $K_2, \Gamma_1, \Gamma_2$ and the CBM $K_2$ we expanded the wave functions of the band states in LCAO-like basis functions. We obtain the projected band structures as shown in Fig.~\ref{fig:transitions}(a) by summing up the squares of the expansion coefficients from all atoms of one layer. The color clearly indicates the localization of the states and allows one to find the energies of the VBM of MoS$_2$ and the CBM of WSe$_2$.
The four most relevant transition energies of HB with different twist angles are given in Fig.~\ref{fig:transitions}(b).
The results show that the ML transitions are insensitive to twisting or interlayer interactions, as their energies are equal to the ML value (dashed lines). The small variations that are discernible are likely to come from differences in the residual strain of the commensurate HB supercells (see Tab.~\ref{tab:HBCells}).
The two interlayer transitions however are sensitive to twisting but only the trend of $\Gamma_1-K_1$ agrees with the PL results, as discussed above and in the main article.

\subsubsection{Comparison with photoluminescence energies}
Absolute DFT transition energies cannot directly be compared with absolute PL energies due to the inability of the single-particle Kohn-Sham description to properly account for both the quasi-particle band gap and the two-particle states such as excitons, that are very important in TMDC materials.
To match the energy scales of PL and DFT in Fig.~{2(d)} of the main article we therefore rigidly shift the DFT transition energies by +0.445 eV. This implies that we do not compare absolute but relative energy scales.
For the variation of the energies with twist angle we obtained quantitative agreement that allows us to ascribe the ILE PL peak to the $\Gamma-K$ transition.
The change of the $\Gamma-K$ transition energy is essentially a shift of the $\Gamma$-point valence band energy (white arrow in Fig.~{3(a)} of the main article). It has been previously shown that relative shifts of valence band energies of TMDCs and other systems with respect to structural changes can be calculated reliably with DFT \cite{Yeh2016a_SI,Zhu1989a_SI}.
The outstanding agreement between experiment and DFT in Fig.~{2(d)} of the main article indicates that neither the dielectric screening nor the exciton binding energies (see Sec.~\ref{sec:excitons}) are significantly affected by twisting and therefore the observed twist-angle-dependent PL shifts are mainly a band structure effect.

\subsubsection{Band alignment diagram: estimation of band gaps and band offsets} \label{sec:BandOffsets}
\begin{figure}[bt]
\begin{center}
\includegraphics[width=.7\textwidth]{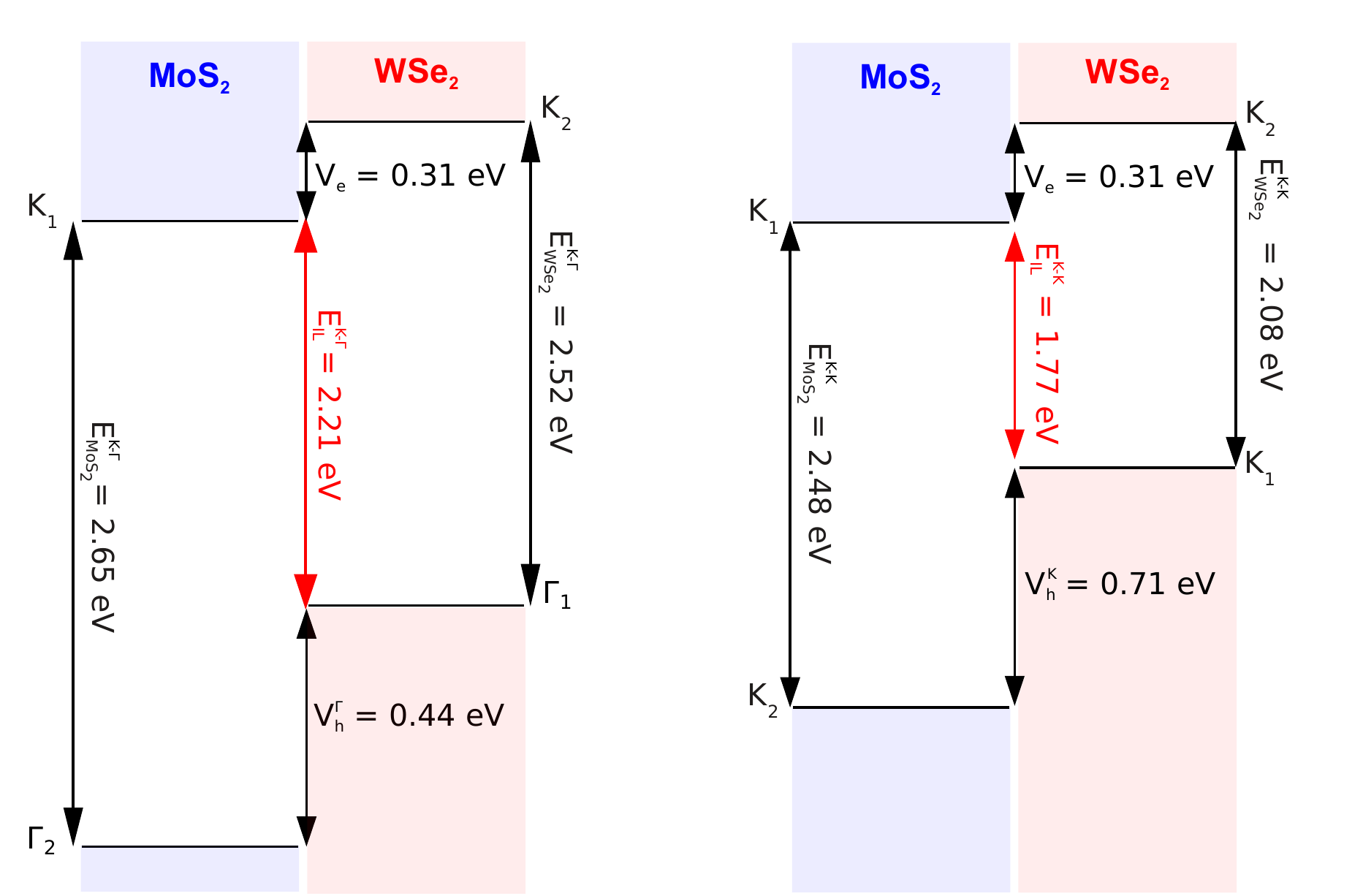}
\end{center}
\caption{\label{fig:BandAlign}
Energy gaps and band offsets in the MoS$_2$/WSe$_2$ heterobilayer for (left) $\Gamma-K$ and (right) $K-K$ transitions. States are labeled as in Fig.~\ref{fig:2H-sep}(a).
}
\end{figure}

The basic optical and electronic properties of a HB are defined by the band edge states. Their relative energies are given in the band alignment diagram in {Fig.3(a)} of the main article and in Fig.~\ref{fig:BandAlign}.
These diagrams contain the band gaps and the band offsets in the MoS$_2$/WSe$_2$ HB.
DFT calculations allow one to calculate ground state properties with high accuracy but band gaps and the band offsets in heterostructures are only qualitatively correct. For a more reliable determination we use converged G$_0$W$_0$ results from Rassmussen et al., who tabulated absolute energies\footnote{relative to the vacuum level} of band edge states ($K_1$ and $K_2$) in MoS$_2$ and WSe$_2$ ML (and various other TMDC{s}) \cite{Rasmussen2015_SI}.
The analysis above shows that interlayer interactions have a weak impact on the band energy of  $K_1$ and $K_2$. Therefore we use these ML values to estimate the band offsets $V_\mathrm{h}^\mathrm{K}$ and $V_\mathrm{e}$ in the HB as given in the right-hand part of Fig.~\ref{fig:BandAlign}.

The band energies of the states $\Gamma_1$ and $\Gamma_2$ are strongly influenced by interlayer interactions (see pink arrows in Fig.~\ref{fig:2H-basic}(b)). As G$_0$W$_0$ calculations for twisted HB (ca.~600 atoms per unit cell) are not at all feasible, we estimate the band  energy of these state as:
$\varepsilon_{\Gamma_1} = \varepsilon_{K_1}^\mathrm{GW} - \frac{1}{2} \Delta_\mathrm{SOC}^\mathrm{GW}(\mathrm{WSe}_2) - \varepsilon_{K_1}^\mathrm{DFT} + \varepsilon_{\Gamma_1}^\mathrm{DFT}$,
where $\varepsilon_{K_1}^\mathrm{GW}$ is the G$_0$W$_0$ band energy,  $\Delta_\mathrm{SOC}^\mathrm{GW}(\mathrm{WSe}_2)$ is the splitting of the $K_1$ state due to spin-orbit interactions from {Ref.}~\cite{Rasmussen2015_SI} and $\varepsilon_{K_1}^\mathrm{DFT}$ and $\varepsilon_{\Gamma_1}^\mathrm{DFT}$ are the DFT band energies in a twisted HB with a twist angle of 21.1\degree (see labels in Fig.~\ref{fig:transitions}(a)). As the  G$_0$W$_0$ calculations include spin-orbit interactions and the DFT calculations do not, we use half of $\Delta_\mathrm{SOC}^\mathrm{GW}(\mathrm{WSe}_2)$ to correct that difference. We argue that it is reasonable to calculate the energy difference between $K_1$ and $\Gamma_1$ in DFT because the two states are mostly localized in the WSe$_2$ layer and DFT band energy differences between valence states in TMDC MLs were shown to be in good agreement with experiment \cite{Jin2013_SI,Yeh2016a_SI}.
Similarly, the band energy of $\Gamma_2$ is estimated as:
$\varepsilon_{\Gamma_2} = \varepsilon_{K_2}^\mathrm{GW} - \frac{1}{2} \Delta_\mathrm{SOC}^\mathrm{GW}(\mathrm{MoS}_2) - \varepsilon_{K_2}^\mathrm{DFT} + \varepsilon_{\Gamma_2}^\mathrm{DFT}$.
The left-hand side of Fig.~\ref{fig:BandAlign} is obtained using this approach.
The band offset of the $\Gamma$-point hole states $V_\mathrm{h}^\mathrm{\Gamma} = \varepsilon_{\Gamma_1} - \varepsilon_{\Gamma_2}$ in separated ML is 0.16 eV and 0.44 eV in HB (21.1\degree).

\subsection{Exciton binding energies}	\label{sec:excitons}

The electron and hole probability density distributions in the MoS$_2$ and WSe$_2$ layers that compose the heterostructure play a fundamental role in the exciton binding energy and oscillator strength. Such distributions can be retrieved from DFT calculations of the band structure of the MoS$_2$/WSe$_2$ heterostructure, as shown in Fig. \ref{fig:2H-basic}(c), where one observes that at the $K$-point, the valence and conduction band states are fully separated in different layers. In fact, this is what one would expect bringing together these two layers: for a type-II band alignment electrons (holes) should be fully confined in the MoS$_2$ (WSe$_2$) layer. This is however not true for the valence band states at the $\Gamma$-point, where a mixing of ML states is observed, with 24$\%$ of the valence orbital  localized in the MoS$_2$ layer.
Whenever we speak of electron-hole overlap $o$ in the figure and discussion below, we refer to the projection{\footnote{
With the resolution of identity $\hat{1} = |\text{WSe}_2\rangle \langle \text{WSe}_2| + |\text{MoS}_2\rangle \langle \text{MoS}_2|$.} of the hole wave function $|+ k \rangle$ ($k=\Gamma$ or $K$) onto the MoS$_2$ layer $o= |\langle \text{MoS}_2|+ k \rangle|^2$.

As a consequence of this mixture, holes at $\Gamma$ and electrons at $K$ points form partially charge-separated excitons in real space that exhibit finite oscillator strength, in contrast to the fully charge-separated exciton, with both electrons and holes at $K$, where these charges are completely spatially separated in different layers and, therefore, have very small oscillator strengths.

In order to estimate the binding energy of such partially charge-separated ILE, one could use the 24$\%$-76$\%$ distribution for the hole, along with a completely localized electron in MoS$_2$, to construct the exciton wave function distribution along the direction perpendicular to the layers, whereas the in-plane motion would be described, e.g., by a variational function. However, electron-hole interactions in 2D materials are known to be very strong, due to the lack of dielectric screening by the environment surrounding the layers (assumed to be vacuum here), which suggests that this distribution, obtained \textit{ab initio} without accounting for electron-hole interactions, may change if this interaction is taken into account. Therefore, it is important to develop a model that matches the DFT {interlayer} orbital distributions in the limit of non-interacting charges, but also allows for corrections due to electron-hole interactions. This is done in what follows, within the tight-binding model framework.

First, we use the DFT results to parametrize the {interlayer} hoppings. By comparing the conduction and valence bands at the $K$-point in Fig.~\ref{fig:2H-basic}(b) for isolated monolayers (dashed lines) and their heterostructure (solid line), one observes that the latter is basically just a superposition of the results in the former. Moreover, as previously mentioned, electrons and holes are fully confined to their  layers, with no {interlayer} mixing. This means that within the simplest tight-binding model for electrons and holes at $K$, {interlayer} hopping parameters must be zero and a basis with electron and hole wave functions given by single-layer states, $|i\rangle$, representing an electron or hole confined in the $i$-th  layer, with $i= $ 1 (for MoS$_2$) or 2 (for WSe$_2$), diagonalizes the system. For such {a} charge-separated exciton, therefore, no further refinement is needed: tight-binding 2$\times$2 matrices for the conduction or valence band Hamiltonian are
diagonal, eigenvalues are
simply 0 and $V^{K}_{e(h)}$ for conduction (valence) bands, where $V^{K}_{e(h)}$ is the conduction (valence) band offset at $K$, and eigenvectors are (1 0) and (0 1). Lowest-energy electron and hole states are (1 0) and (0 1), thus representing an electron localized in MoS$_2$ and a hole in WSe$_2$, and the binding energy can be readily calculated just by solving the in-plane Schr\"odinger equation for an electron-hole pair using the interaction potential for separate carriers, which will be discussed in greater {detail} further on.

For valence band states at $\Gamma$ in Fig. \ref{fig:2H-basic}(c), however, there is a clear mixture of ML states and the heterostructure valence band is not just a superposition of the monolayer ones. This suggests that the 2 $\times$ 2 tight-binding Hamiltonian matrix in this case is no longer diagonal, but, in the simplest tight-binding model, can be rather written as
\begin{equation}
H_v^{\Gamma} =  \left(
\begin{tabular}{cc} $V^{\Gamma}_{h}$ & $t_h$\\
$t_h$ & $0$
\end{tabular}
\right),
\end{equation}
where $V^{\Gamma}_{h}$ is the valence band offset between separate ML and $t_h$ is the {interlayer} hopping parameter.
The band structure in Fig. \ref{fig:2H-basic}(b) shows that the band offset (pink arrows) increases as the ML are put together in a HB.
In the previous section $V^{\Gamma}_{h}$ was determined to be 0.16 eV and 0.44 eV for separated ML and the HB, respectively.
We thus adjust $t_h$ as to reproduce this increase of the band offset in the heterostructure valence bands, obtaining $t_h = 0.2341$ eV. Finally, diagonalizing $H_v^{\Gamma}$ for these parameters, one obtains also a hole wave function that is distributed between both layers, with 32$\%$(68$\%$) probability to be found in the MoS$_2$ (WSe$_2$) - very close to the ratio 24$\%$-76$\%$ obtained with DFT, which adds support to our model.

We now consider a hybrid trial wave-function for the exciton $|\Psi\rangle = H_y(r)|ij\rangle$: its radial part is given by a variational hydrogenic function $H_y(r) = (2\big/a\sqrt{2 \pi}) e^{-r/a}$, where $a$ is a variational parameter, whereas its longitudinal part $|ij\rangle$ consists of the tight-binding basis discussed above, with an electron in the $i$-th layer and a hole in the $j$-th layer. In this way, we are assuming that the wave function spreads over the planes of each layer (with an effective Bohr radius $a$), but it is infinitely thin around each layer in the direction perpendicular to them. This approach allows us to ''lock'' the eigenstates in the ground state for the radial direction, while we investigate the electron-hole distribution along the layers for different states in the perpendicular direction.

\begin{figure}[!b]
\centering
\includegraphics[width=.5\textwidth]{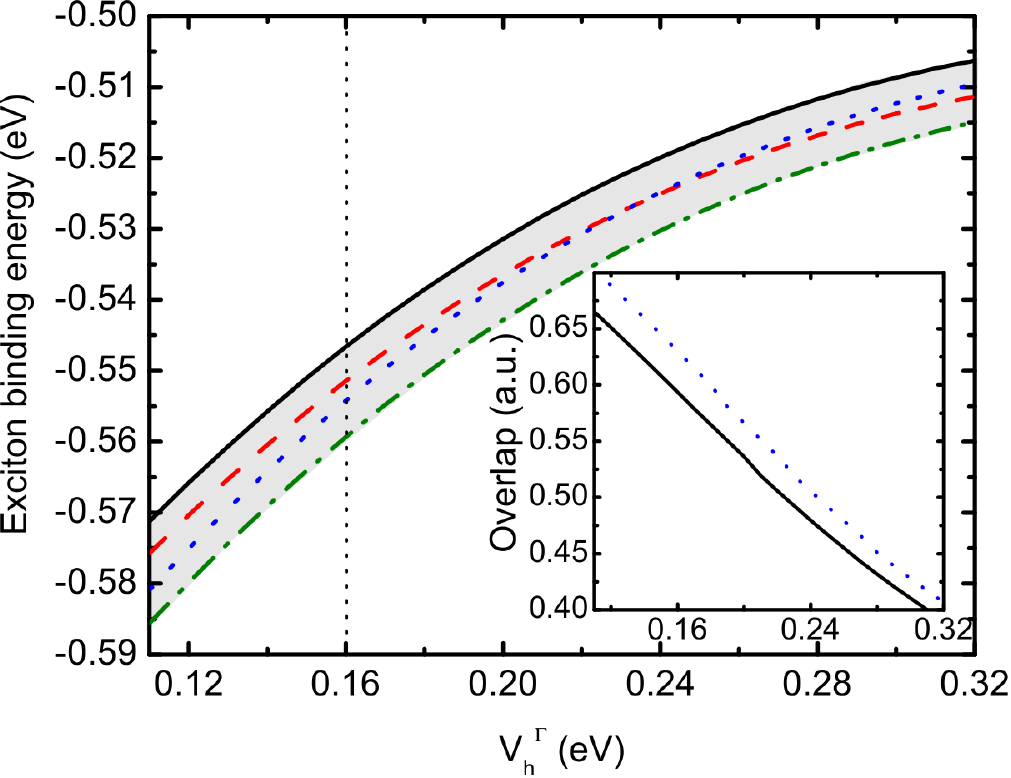}
\caption{\label{fig:excitonenergy} Ground state K-$\Gamma$ ILE binding energy as a function of the hole band offset, for different  hole effective masses:
$m^{\Gamma,WSe_2}_h = 2.1$ m$_0$ and $m^{\Gamma,MoS_2}_h = 3.3$ m$_0$ (black solid), $m^{\Gamma,WSe_2}_h = 2.1$ m$_0$ and $m^{\Gamma,MoS_2}_h = 10.0$ (blue dotted), $m^{\Gamma,WSe_2}_h = 3.7$ m$_0$ and $m^{\Gamma,MoS_2}_h = 3.3$ m$_0$ (red dashed), and $m^{\Gamma,WSe_2}_h = 3.7$ m$_0$ and $m^{\Gamma,MoS_2}_h = 10.0$ (green dashed-dotted). Electron-hole overlaps are shown in the inset for the black solid and blue dotted cases. {The abscissa of the inset agrees with the main plot.}
}
\end{figure}

In the model basis $|11 \rangle$, $|12 \rangle$, $|21 \rangle$, $|22 \rangle$, the hybrid continuum/tight-binding model Hamiltonian is represented by the matrix
\begin{equation}\label{eq.Ham}
H =  \left(
\begin{tabular}{cccc} $H_{11,11}$ & $t_h$ & $t_e$ & 0\\
$t_h$ & $H_{12,12}$ &  0 & $t_e$\\
$t_e$ & 0 & $H_{21,21}$ & $t_h$ \\
$0$ & $t_e$ & $t_h$ & $H_{22,22}$
\end{tabular}
\right),
\end{equation}
with diagonal elements $H_{11,11} = K_{11}+V_{11}+V^{\Gamma}_h$, $H_{12,12} = K_{12}+V_{12}$, $H_{21,21} = K_{21}+V_{21} + V^{K}_e + V^{\Gamma}_h$, and $H_{22,22} = K_{22}+V_{22} + V^K_e$, where the kinetic energy contribution is
\begin{equation}
K_{ij} = \frac{\hbar^2}{2\mu_{ij}}\frac{1}{a^2},
\end{equation}
with reduced effective masss $\mu_{ij} = (1/m^i_e + 1/m^j_h)^{-1}$, while the effective {inter- and intralayer} electron-hole interaction potentials $V_{ij}$ are the expectation values of the interaction potential $V^{QEH}_{ij}(r)$, calculated within the Quantum Electrostatic Heterostructure model \cite{Andersen2015_SI}, assuming an electron in the $i$-th layer and a hole in the $j$-th layer, i.e. $V_{ij} = 2\pi\int_{0}^{\infty}H_{y}^2(r)V^{QEH}_{ij}(r)rdr$. Inter-layer hopping parameters for electrons and holes are $t_{e}$ and $t_h$, respectively. The Hamiltonian matrix Eq. (\ref{eq.Ham}) is then numerically diagonalized for different values of $a$, until a minimum ground state energy is reached.

As verified in Fig. \ref{fig:BandAlign}, the conduction band offset at $K$ is 0.31 eV, (MoS$_2$ being the layer with lowest energy) and due to the lack of {interlayer}  mixing of the conduction band orbitals, we assume $t_e = 0$, which makes the HB conduction bands the same as a superposition of ML bands, as previously discussed. This is equivalent to having a very high electron effective mass in $z$-direction for the TMDC layers. The in-plane electron masses are $m^{MoS_2}_e= 0.55$ m$_0$ and $m^{WSe_2}_e = 0.48$ m$_0$ \cite{Rasmussen2015_SI}.  For the valence band we take $V^{\Gamma}_h = 0.16$ eV, $t_h = 0.2341$ eV and the effective masses $m^{\Gamma,MoS_2}_h = 3.3$ m$_0$ and $m^{\Gamma,WSe_2}_h = 2.1$ m$_0$ \cite{Peelaers2012_SI}, unless otherwise explicitly stated.

Let us first investigate the A exciton binding energies in isolated ML and in the HB within our model. In this case,  the variational part of the model is sufficient and the tight-binding matrix construction is discarded, since electrons and holes are both localized in a single {layer and} involve $K$-point states only.
The A exciton binding energies are thus obtained simply by minimization of $E^{MoS_2(WSe_2)}_b = K_{11(22)} + V_{11(22)}$ with respect to $a$. For isolated ML we obtain $E^{MoS_2}_b = -0.607$ eV and $E^{WSe_2}_b = -0.553$ eV. These values are in good agreement with those in the literature \cite{Berkelbach2013_SI, GangWang_SI}.
For the MLs in the HB we obtain $E^{MoS_2}_b = -0.432$ eV and $E^{WSe_2}_b = -0.396$ eV , in excellent agreement with results of Latini et al.~who obtained 0.42 eV and 0.39 eV, respectively  \cite{Latini2017_SI}.
These values are lower than those obtained for isolated ML. This is due to the fact that in the HB, even when dealing with ML excitons, $V_{ii}$ is calculated taking into account the existence of an adjacent layer, which provides extra screening of the Coulomb interaction thus reducing ML binding energies.

The $K-K$ ILE does not require the use of the tight-binding matrix described here since the conduction and valence band states in $K$ are contained in each single layer. Equivalently, the tight-binding model developed here would have hopping parameters $t_e = t_h = 0$. The hole effective masses at $K$ are $m^{K,MoS_2}_h = 0.56$ m$_0$ and $m^{K,WSe_2}_h = 0.44$ m$_0$ \cite{Rasmussen2015_SI}.
Our model leads to a full $K-K$ charge-separated exciton with binding energy $E^{K-K}_b = -0.285$ eV. This result is in good agreement with $0.28$ eV found in Ref. \cite{Latini2017_SI}, where only the $K-K$ ILE is discussed.

Conversely, for the partially charge-separated $K-\Gamma$ ILE, diagonalization of the Hamiltonian in Eq.~\ref{eq.Ham} is required, which leads to a binding energy $E^{K - \Gamma}_b = -0.547$ eV , with {an electron-hole overlap of 0.59}.
As the {interlayer} gap for the $K - \Gamma$ transition is 2.21 eV (see Fig. \ref{fig:BandAlign}), the ILE peak for this transition is expected to be at $E^{K-\Gamma}_{ILE} - E^{K-\Gamma}_b = $ 1.663 eV. Notice the {interlayer} electron-hole attraction is so strong it changes the hole distribution between layers from the previous 32$\%$-68$\%$, for the non-interacting case, to 59$\%$-41$\%$ when the interaction is turned on. Such a hole distribution, strongly shared between both layers,  explains the high values found for binding energy and electron-hole overlap in this case.

It is worth investigating how these results depend on the effective mass and band offsets chosen in the model. For example, as discussed in Sec. \ref{sec:BandOffsets}, the $V^{\Gamma}_h = 0.16$ eV band offset used here is inferred from the one found with G$_0$W$_0$ calculations for the $K-K$ transition by subtracting the  difference between the top of the valence bands for each material at $\Gamma$ and $K$ points as obtained from DFT. There may thus be a possible inaccuracy in the definition of this parameter which requires further investigation of the ILE dependence on this parameter. For the effective masses previously mentioned, the dependence of the ILE binding energy $E^{K - \Gamma}_b$ on $V^{\Gamma}_h$ is shown as a solid black curve in Fig. \ref{fig:excitonenergy}. One verifies that even doubling the band offset to $V^{\Gamma}_h = 0.32$ eV, the binding energy still decreases only by $0.04$ eV, less than $10 \%$ of its original value at $V^{\Gamma}_h = 0.16$ eV. Notice the value $m^{\Gamma,WSe_2}_h
= 2.1$ m$_0$ is obtained from the heterostructure band structure, whereas $m^{\Gamma,MoS_2}_h = 3.3$ m$_0$ is the monolayer effective mass for MoS$_2$ (at $\Gamma$). In fact in the heterostructure, the valence band whose wave function is predominantly localized in the MoS$_2$ ML is almost flat [see Figs. \ref{fig:2H-basic}(b) and \ref{fig:transitions}(a)], i.e. it exhibits an extremely high effective mass. Nevertheless, in the definition of the reduced mass $\mu_{ij}$, this mass appears in the denominator, so that high values just shift the reduced mass closer to the electron mass and the binding energy {would  not be} significantly affected. This is verified by the blue dotted curve in Fig. \ref{fig:excitonenergy}, which shows results assuming $m^{\Gamma,MoS_2}_h = 10.0$ m$_0$. Such an enhancement of the hole effective mass for the $\Gamma_2$ band only leads to {a~$\lesssim$~0.010~eV} difference in the ILE binding energy, for any value of $V^{\Gamma}_h$. The hole effective mass for the $\Gamma_1$
band, on the other
hand, is reduced  in the heterostructure.
Using the ML value $m^{\Gamma,WSe_2}_h = 3.7$ m$_0$ while keeping $m^{\Gamma,MoS_2}_h = 3.3$ m$_0$, one obtains the red dashed curve in Fig. \ref{fig:excitonenergy}, which is $\lesssim$ 0.007 eV below the black solid curve. Finally, using extreme values of effective masses, i.e. $m^{\Gamma,WSe_2}_h = 3.7$ m$_0$ and $m^{\Gamma,MoS_2}_h = 10.0$, still leads to binding energies only $\approx$ 0.010 eV below the black solid curve, as shown by the green dashed-dotted curve in Fig. \ref{fig:excitonenergy}. Intermediate values of effective mass would lead to curves within the shaded region of this figure. It is thus safe to say that the ILE binding energy is only weakly dependent on the effective masses, which is in agreement with discussions in previous theoretical works \cite{Olsen2016_SI}. Although the dependence on $V^{\Gamma}_h$ is stronger, changing this parameter still does not lead to values of ILE binding energy far
from the {$\approx 0.5$} eV found here and the conclusion that the ILE has high binding energy is robust.

The influence of $V^{\Gamma}_h$ on the electron-hole overlap $o$
is shown in the inset of Fig. \ref{fig:excitonenergy} for the cases where $m^{\Gamma,WSe_2}_h = 2.1 $ m$_0$ and $m^{\Gamma,MoS_2}_h = 3.3$ m$_0$ (black solid) or $m^{\Gamma,MoS_2}_h = 10.0$ m$_0$ (blue dotted). This overlap is closely related to the interlayer dipole moment ($\mu_\mathrm{IL} = ed(1-o)$, $e$ and $d$ are the elementary charge and the layer separation, respectively), which could be experimentally verified by measuring the ILE Stark shift due to a perpendicularly applied electric field, where a linear contribution should appear along with the usual quadratic shift. This measurement would thus give also an estimate of the hole band offset based on the model proposed here.

Let us now compare the theoretically obtained results with the actual peaks observed in the PL experiment. The mean values of the A {exciton} peaks of MoS$_2$ and WSe$_2$ ML are 1.920 eV {(here, we assume an exciton-trion splitting of 30~meV)} and  1.659 eV, respectively. Using the energy gap at $K$-point for MoS$_2$ and WSe$_2$ in Fig.~\ref{fig:BandAlign} and the previously calculated exciton binding energies for isolated ML, these peaks are theoretically predicted to be around 1.873 and 1.527 eV, respectively. Thus we obtain excellent agreement for MoS$_2$ and a small underestimation of the PL energy for WSe$_2$.
The ILE peak in the HB is measured at $\approx 1.61$ eV (twist angle near 30 degrees). Our theoretical value for the $\Gamma-K$ ILE  is 1.663 eV. This value lies within the same order of magnitude as the ML WSe$_2$ peak. This explains why the lower energy peaks in the experimental PL spectrum are practically merged.
Moreover, using the {interlayer} $K-K$ gap $E^{K-K}_{IL}$ from Fig. \ref{fig:BandAlign}, combined with the binding energy for the $K-K$ ILE, $E^{K-K}_b$, leads to a peak around 1.485 eV, which is already far from the experimentally observed merged peaks around $\approx 1.60-1.65$. This provides further support that not a $K-K$ ILE but  $\Gamma-K$ ILE is experimentally observed.

%

\end{document}